\documentclass[3p]{elsarticle}

\graphicspath{{./figures/}}
\usepackage{subcaption}
\usepackage[cmex10]{amsmath}
\usepackage{amsfonts}
\usepackage{algorithm}
\usepackage{algorithmic}
\usepackage{color}
\usepackage[]{amsthm}
\usepackage{setspace,caption,url}
\usepackage{array}
\newtheorem{theorem}{Theorem}
\newcommand{\der}{{d}}
\newcommand{\bc}{\mathbb{C}}
\newcommand{\thetabar}{\bar{\theta}}
\newcommand{\Thetabar}{\bar{\Theta}}
\newcommand{\hp}{\widehat{\theta}}
\newcommand{\hS}{\widehat{S}}
\newcommand{\hs}{\widehat{s}}
\newcommand{\hc}{\widehat{c}}

\newcommand{\dw}{\,\der  \omega}

\newcommand{\TDE}{TDE}
\newcommand{\FE}{FE}
\newcommand{\CS}{CS}
\newcommand{\CPE}{compressive parameter estimation}
\newcommand{\PD}{PD}
\newcommand{\PDs}{PDs}
\newcommand{\EMD}{EMD}
\newcommand{\PEE}{parameter estimation error}

\begin{document}

\begin{frontmatter}

\title{Performance of Compressive Parameter Estimation \\ via K-Median Clustering \tnoteref{titleNote}}
\tnotetext[titleNote]{Early versions of this work appeared at Proceedings of SPIE Wavelets and
Sparsity XV, August 2013 \cite{mo2013compressive} and 2016 Annual Conference on Information Science
and Systems\cite{Mo2016Compressive-Lin}}

\author[mainAddress]{Dian Mo}
\ead{mo@umass.edu}
\author[mainAddress]{Marco F. Duarte}
\ead{mduarte@ecs.umass.edu}
\address[mainAddress]{Department of Electrical and Computer Engineering, University of Massachusetts, Amherst, MA 01003}

\begin{abstract}
Compressive sensing (\CS) has attracted significant attention in parameter estimation tasks, where
parametric dictionaries (\PD s) collect signal observations for a sampling of the parameter space
and yield sparse representations for signals of interest when the sampling is dense. While this
sampling also leads to high dictionary coherence, one can leverage structured sparsity models to
prevent highly coherent dictionary elements from appearing simultaneously in the recovered signal.
However, the resulting approaches depend heavily on the careful setting of the maximum allowable
coherence; furthermore, their guarantees are not concerned with general parameter estimation
performance. We propose the use of earth mover's distance (\EMD), as applied to a pair of true and
estimated \PD~coefficient vectors, to measure the parameter estimation error. We formally analyze
the connection between the \EMD~and the parameter estimation error and show that the \EMD~provides a
better-suited metric for parameter estimation performance than the Euclidean distance. Additionally,
we analyze the previously described relationship between $K$-median clustering and \EMD-optimal
sparse approximation and leverage it to develop improved PD-based parameter estimation algorithms.
Finally, our numerical experiments verify our theoretical results and show that the proposed
compressive parameter estimation algorithms have similar performance as the state-of-the-art
algorithms while featuring simpler implementation and broader applicability.

\end{abstract}

\begin{keyword}
compressive sensing; parameter estimation; parametric dictionary; earth mover's distance; $K$-median
clustering
\end{keyword}

\end{frontmatter}

\section{Introduction}


Compressive sensing (\CS) simultaneously acquires and compresses signals via random projections, and
recovers a signal if it is sparse or if there exists a basis or dictionary in which it can be expressed
sparsely \cite{baraniuk2007compressive, candes2006compressive, donoho2006compressed}.  Recently, the
application of \CS~has been extended from signal recovery to parameter estimation through the design
of parametric dictionaries (\PD s) that contain signal observations for a sampling set of the
parameter space \cite{cevher2008compressive, duarte2012localization, herman2009high,
jamali-rad2013sparsity-aware, sen2011multiobjective, stojanovic2009compressed, eftekhari2013matched,
fyhn2013compressive, austin2010on-the-relation, bourguignon2007a-sparsity-based, duarte2013spectral,
fyhn2013spectral}. The resulting connection between parameter estimation and sparse signal recovery
has made it possible for compressive parameter estimation to be implemented via standard
\CS~recovery algorithms, where the \PD~coefficients obtained from recovery algorithms can be
interpreted by matching the locations of the nonzero coefficients with the parameter estimates. This
\CS~approach has been previously formulated for many landmark \CPE~problems, including localization
and bearing estimation \cite{cevher2008compressive, duarte2012localization, herman2009high,
jamali-rad2013sparsity-aware, sen2011multiobjective, stojanovic2009compressed}, time delay
estimation \cite{eftekhari2013matched, fyhn2013compressive}, and frequency estimation (also known as
line spectral estimation when the frequencies lie in a one-dimensional space)
\cite{austin2010on-the-relation, bourguignon2007a-sparsity-based, duarte2013spectral,
fyhn2013spectral, fannjiang2012coherence}.

Unfortunately, only in the contrived case when the unknown parameters are all contained in the
sampling set of the parameter space can the PD-based parameter estimation be perfect. In frequency
estimation, the mismatch between the unknown frequencies and the discretized frequency samples
results in loss of sparsity due to spectral leakage, therefore reducing recovery performance
\cite{Chi2011Sensitivity-to-}. Additionally, the guarantees of bounded error between the true and
the estimated \PD~coefficient vectors, measured via the Euclidean distance in almost all existing
methods, have very limited impact on the performance of \CPE~given that the Euclidean distance does not
relate to the parameter estimation error. In contrast, the earth mover's distance (\EMD)
\cite{rubner1998a-metric, gupta2010sparse, indyk2011k-median} is a very attractive option due to the
fact that the \EMD~of two sparse \PD~coefficient vectors is indicative of the parameter estimation
error when the elements of the \PD~are sorted in increasing order of the values of the corresponding
parameters.

\subsection{Prior Work}

The relevant literature includes several approaches that rely on convex optimization to avoid 
parameter discretization issues.  Recently, total variation norm minimization has been proposed 
to estimate frequency-sparse signals from observations whose spectrum is confined within some 
set of low frequencies, a setup that is now known as the super-resolution problem
\cite{Candes2014Towards-a-Mathe}. This algorithm allows for accurate frequency estimation 
when the unknown frequencies are sufficiently separated, even under additive noise
\cite{candes2013super-resolution}. Inspired by this approach, when the frequencies are well
separated, atomic norm minimization has been developed to estimate frequency-sparse signals 
with line spectra from samples obtained at 
random times~\cite{bhaskar2013atomic,tang2013compressed}.  An
extension to two-dimensional frequency estimation is available in \cite{Chi2013Compressive-Rec}.
Another method uses matrix completion to recover signals with arbitrary frequencies from random
samples when a separation condition is met \cite{Chen2011Robust-Spectral}. The main result is 
that the sparse recovery problem can be solved via low-rank matrix completion of a matrix 
possessing Hankel structure derived from the sparse signal. Similar performance guarantees have 
been established as well for signal recovery from linear 
measurements~\cite{Cai2015Robust-Recovery}. The main drawback of optimization methods 
is that the performance is highly dependent on convex optimization solvers, which usually can 
be very time-consuming. Additionally, such methods are hard to extend to other parameter 
estimation problems such as time delay estimation.

For \PD-based parameter estimation, a dense sampling of the parameter space is needed to 
improve the estimation resolution \cite{duarte2012localization}. However, increasing the 
sampling density introduces highly coherent elements in the \PD s, which might affect the 
performance and complexity of sparse recovery algorithms. Previous approaches to address the 
resulting coherence issues utilized a coherence-inhibiting structured sparse approximation where 
the resulting nonzero entries of the reconstructed coefficient vector correspond to the 
\PD~elements that have sufficiently low 
coherence~\cite{fannjiang2012coherence, duarte2012localization, duarte2013spectral}. 
Based on structured sparse recovery, an auxiliary interpolation has been proposed as an 
additional step to break through the limitation of discretization and improve the estimation 
performance~\cite{fyhn2013compressive, fyhn2013spectral}. Those approaches need a 
careful setting of the value of the maximum allowable coherence among the chosen 
\PD~elements in the sparse representation. Setting the maximum allowable coherence to be 
too large restricts the minimum separation between any two parameters that can be observed 
simultaneously in a signal, and can potentially exclude a large class of observable signals from 
consideration.

\EMD~has been proposed as an alternative measure for sparse recovery \cite{indyk2011k-median}.
Instead of performing constrained \EMD~minimization directly, a pyramid transform that connects a
sparse model and a tree-sparse model allows for the implementation of sparse recovery with error
measured by \EMD~using a constrained $\ell _1$-norm minimization problem. However, the special
requirements of the measurement matrix limit the use of such \EMD-optimal sparse recovery in many
practical applications.

\subsection{Contribution}

In this paper, we proposed a new \PD-based method for \CPE~that
replaces the hard thresholding operator, a widely used sparse approximation operator in sparse
recovery algorithms to solve the sparse approximation problem under Euclidean distance, with
$K$-median clustering, a novel sparse approximation operator that returns the optimal sparse
approximation vector in terms of the \EMD~\cite{gupta2010sparse, indyk2011k-median}. Our main
contributions can be detailed as follows.  First, we show theoretically that the \EMD~between the
sparse \PD~coefficient vectors provides an upper bound for the parameter estimation error,
motivating the use of algorithms minimizing the \EMD~between coefficient vectors to achieve small
parameter estimation error. Second, we formulate theorems that provide performance guarantees for
parameter estimation with our proposed clustering methods; these guarantees are based on the
correlation between \PD~elements.  Third, we analyze the effect of the decay of the correlation
between \PD~elements on the performance of our proposed method, and elaborate on our guarantees when
\CS~compression and signal noise are involved. Finally, we introduce and analyze the joint use of
thresholding and clustering methods to address performance loss resulting from compression and
noise. Although this paper focuses on one-dimensional parameters, our work can be easily extended to
multidimensional parameter estimation by using $K$-median clustering for multidimensional spaces.

One of the essential advantages of our proposed algorithm is that it can be applied to arbitrary
sparsity-based parameter estimation problems. As shown in the sequel, existing greedy algorithms
require an additional parameter to prevent highly coherent \PD~elements from appearing in the signal
representation simultaneously and estimate the correct parameters. As shown in our prior work, such
algorithms are very sensitive to the chosen value of the additional parameter: an improper value of
the additional parameter is significantly harmful to the estimation performance
\cite{mo2013compressive}. However, to the best of our knowledge, there is no guarantee that existing
greedy algorithms with a specific parameter value will have good performance for parameter
estimation under suitable conditions.  Furthermore, we do not know of a procedure to set the
parameter values so that the existing algorithms will have good performance under the given
conditions except for a grid search that chooses the one that performs best experimentally. In
contrast, our proposed method does not rely on additional parameters in the most amenable cases. In
the presence of a slowly decaying correlation function, compression, or signal noise, where it is
more difficult for our proposed algorithm to obtain accurate estimates, our theoretical analysis
provides guidelines to choose a proper value of the threshold level and guarantees that the proposed
algorithm will have small estimation error. In terms of broad applicability, our proposed method
can be used for time delay estimation, frequency estimation and other sparsity-based parameter
estimation problems; additionally, the observation model is arbitrary as well, with example
applications using subsampled signals or randomly projected measurements. This is in contrast
with existing approaches that obtain state-of-the-art performance, which are specialized to a single
problem and observation model.

\subsection{Organization}

This paper is organized as follows. In Section \ref{section:background}, we provide a summary of
\CPE~and the issues in existing work and introduce some important background for the proposed
algorithm. In Section~\ref{section:cpe}, we present and analyze the use of \EMD~and clustering
methods for \PD-based parameter estimation; furthermore, we formulate and analyze an algorithm for
\EMD-optimal sparse recovery that employs $K$-median clustering. In
Section~\ref{section:experiments}, we present numerical simulations that verify our results for the
clustering method in the example applications of time delay estimation and frequency estimation.
Finally, we provide a discussion and conclusions in Section~\ref{section:conclusion}.

\section{Background}
\label{section:background}

\subsection{One-Dimensional Parameter Estimation}

One-dimensional parameter estimation problems are usually defined in terms of a parametric signal
class, which is defined via a mapping $\psi : \Theta \rightarrow X$ from the parameter space $\Theta
\subset \mathbb{R}$ to the signal space $X \subset \mathbb{C}^N$. The signal observed in a parameter
estimation problem contains $K$ unknown parametric components $x = \sum_{i=1}^K c_i\psi(\theta_i)$,
and the goal is to obtain estimates of the parameters $\widehat{\theta}_i$ from the signal $x$.  For
example, in time delay estimation, the parameter $\theta_i = \tau _i$ refers to the time delay and the
parametric signal $\psi(\theta_i)$ corresponds to a known waveform $g(t)$ with time delay $\tau_i$,
i.e.,
\begin{align}
    \psi(\theta_i) = \left[g(- \tau_i), g(T - \tau _i), \dots, g((N-1)T - \tau _i)\right]^T, 
\end{align}
when $T$
represents the sampling period. Similarly, in line spectral estimation the parameter $\theta_i = f_i$
is the frequency and the parametric signal corresponds to the complex exponential of the
frequency $f_i$, i.e.,
\begin{align}
    \psi(\theta_i) = \mathbf{e}(f_i) = \left[1, e^{j 2 \pi f_i \frac{1}{N}}, \dots, e^{j 2 \pi f_i \frac{N-1}{N}} \right]
    ^{T}.
\end{align}

Due to the demands in storage, communication, and computation, for large-scale signal processing it
is often desirable to work with a lower-dimensional observation $y$ rather than the signal $x$
itself. The most common observations $y = x_S$ contain only the signal samples at random times $S
\subset \{1, 2, \dots, N\}$ with $|S| = M$. Alternatively, the observations $y = \Phi x$ can also be
obtained by compressing the signal with a dimension-reducing measurement matrix $\Phi \in
\mathbb{R}^{M \times N}$.

Although we are not aware of convex optimization approaches for generic time delay estimation, it is
possible to transform a time delay estimation problem into a frequency estimation problem when the
signal observations are linear measurements of the signal \cite{Saariiisaari1997TLS-ESPRIT-in-a,
Karsten-Fyhn2015Compressive-Par}, which can then itself be solved using convex optimization
\cite{bhaskar2013atomic, tang2013compressed, Chi2013Compressive-Rec, Chen2011Robust-Spectral,
Cai2015Robust-Recovery}. Assume that $G$ is a diagonal matrix whose diagonal entries are the
discrete Fourier transform coefficients of the waveform samples $g = [g(0), g(T), \dots,
g((N-1)T)]^T$. Then $X$, the discrete Fourier transform of $x$, has the sparsity structure as $X = G
x'$, where $x'= \sum _{i=1} ^K c_i \mathbf{e}(\tau _i)$ is a frequency-sparse signal. Thus, the
linear measurements can be expressed as
\begin{align}
    y = \Phi x = \Phi F ^{-1} X = \Phi F ^{H} G x'.
    \label{equation:modified}
\end{align}
Therefore, the time delay estimation of $x$ becomes the frequency estimation of $x'$ with the new
measurement matrix $\Phi' = \Phi F^H G$. In summary, to estimate the time delay indirectly via
frequency estimation, the observations are required to be the linear measurements of the signals,
and the diagonal matrix $G$ should have nonzero diagonal entries so that the new measurements
matrix $\Phi'$ satisfies the properties required by sparse recovery algorithms. We will see that
obtaining such $G$ is difficult in practice in Section~\ref{section:experiments}.

\subsection{Parametric Dictionaries}
The parameter estimation methods based on \PD s model the estimation problem in a generic
form and can be used both in frequency estimation and time delay estimation directly. One can obtain
a \PD~as a collection of samples from the parametric signal space 
\begin{align}
    \Psi = \left[ \psi \left( \thetabar_1 \right), \psi \left( \thetabar_2 \right), \dots, \psi \left( \thetabar _L \right) \right] \subseteq \psi ( \Theta ), 
\end{align}
which corresponds to a set of samples from the parameter space $\Thetabar = \left\{ \thetabar _1,
\thetabar _2, \dots, \thetabar _L \right\} \subseteq \Theta$. In this way, the signal can be
expressed as a linear combination of the \PD~elements $x = \Psi c$ when all the unknown parameters
are contained in the sampling set of the parameter space, i.e., $\theta_i \in \Thetabar$ for each $i
= 1,\ldots,K$. Therefore, finding the unknown parameters reduces to finding the \PD~elements
appearing in the signal representation or, equivalently, finding the nonzero entries or support of
the sparse \PD~coefficient vector $c$ for which we indeed have $y = \Phi\Psi c$. The search for the
vector $c$ can be performed using \CS~recovery.

\PD-based \CPE~can be perfect only if the parameter sample set $\Thetabar$ is dense and large enough
to contain all of the unknown parameters $\{\theta_1, \theta_2, \dots, \theta_K\}$. If this
stringent case is not met for some unknown parameter $\theta_k$, a denser sampling of the parameter
space decreases the difference between the unknown parameter $\theta_k$ and the nearest parameter
sample $\thetabar_l$, so that we can better approximate the parametric signal $\psi(\theta_k)$ with
the parametric signal $\psi \left(\thetabar_l \right)$. However, highly dense sampling increases the
similarity between adjacent \PD~elements and, by extension, the \PD~coherence
\cite{donoho2003optimally}, corresponding to the maximum normalized inner product of \PD~elements:
\begin{align}
\mu(\Psi) = \max_{1 \le i \ne j \le L}  \frac{\left| \langle \psi(\thetabar_i), \psi(\thetabar_j) \rangle
\right|}{ \left\| \psi(\thetabar_i) \right\|_2 \left\| \psi(\thetabar_j) \right\|_2}.
\end{align}
Additionally, denser sampling increases the difficulty of distinguishing between \PD~elements and severely
hampers the performance of \CPE~\cite{tropp2004greed, rauhut2008compressed}. Prior work addressed
such issues by using a coherence-inhibiting structured sparse approximation, where the resulting $K$
nonzero entries of the coefficient vector correspond to \PD~elements that have sufficiently low
coherence. This approach is known as band exclusion \cite{fannjiang2012coherence,
duarte2012localization, fyhn2013compressive, duarte2013spectral, fyhn2013spectral} and inhibits the
highly coherent \PD~elements from appearing in signal representation simultaneously. Choosing a
adequate value for the maximum allowed coherence $\nu$ that defines the restriction on the choice of
\PD~elements is essential to successful performance: setting its value too large results in the
selection of coherent \PD~elements, while setting its value too small tightens up requirements on
the minimum separation of the parameters. Unfortunately, a guideline for the chosen of $\nu$ has not
been fully established.

Another issue is that existing \CS~recovery algorithms commonly used in this setting can only
guarantee stable recovery of the sparse \PD~coefficient vectors when the error is measured by the
$\ell_2$ norm; in other words, we can only predict that the estimated coefficient vector will be
close to the true coefficient vector in Euclidean distance. Such a guarantee is linked to the core
hard thresholding operator, which returns the optimal sparse approximation to the input vector with
respect to the $\ell_2$ norm. However, the guarantee provides control on the performance of
parameter estimation only in the most demanding case of exact recovery, i.e., when the parameter
estimation is perfect.  Otherwise, such a recovery guarantee is meaningless for parameter estimation
since the $\ell_2$ norm cannot precisely measure the difference between the supports of the sparse
\PD~coefficient vectors. For an illustrative example, consider a simple frequency estimation problem
where the PD collects complex sinusoids at frequencies $\{f_1, f_2, \dots, f_L\}$ and there is only
one unknown frequency $f_i$ ($i \in \{1, 2, \dots, L\}$). In other words, the coefficient vector is
$c = e_i$, where $e_i$ denotes the canonical vector that is equal to $1$ at its $i^{\text{th}}$
entries and $0$ elsewhere.  Although two estimated coefficient vectors $\hc_1 = e_j$ and $\hc_2 =
e_k$ ($j, k \in \{1, 2, \dots, L \}$ and $j \ne k$) have the same $\ell_2$ distance to the true
coefficient vector $c$, the frequency estimate $f_j$ from $\hc_1$ and $f_k$ from $\hc_2$ have
unequal estimation error $|f_i - f_j|$ and $|f _i - f _k|$, respectively.

Alternatively, the earth mover's distance (\EMD) has recently been used in \CS~to measure the
distance between coefficient vectors in terms of the similarity between their supports
\cite{gupta2010sparse, indyk2011k-median}. In particular, the \EMD~between two vectors with the same
$\ell_1$ norm optimizes the work of the flow (i.e., the amount of the flow and the distance of flow)
applied to one vector in order to obtain the other vector. In our illustrative example, if the
values of the frequency samples of the PD increase monotonically, the \EMD s between the true
coefficient vector $c$ and the estimated coefficient vectors $\hc_1$ and $\hc_2$ are proportional to
the frequency errors $|f_i - f_j|$ and $|f_i - f_k|$, respectively. Based on the fact that the work
of the flow between any two entries of a PD coefficient vector is proportional to the distance
between the two corresponding parameters, it follows that the \EMD~between pairs of PD coefficient
vectors efficiently measures the corresponding error of parameter estimation. We elaborate our study
of this property in Section~\ref{section:emd}.

\subsection{$K$-Median Clustering}

Cluster analysis partitions a set of data points based on the similarity information between
each pair of points, which is usually expressed in terms of a distance~\cite{bradley1997clustering,
tan2005introduction}. Clustering is the task of partitioning a set of points into different groups
in such a way that the points in the same group, which is called a cluster, are more similar to each
other than to those in other groups. The greater the similarity within a group is and the
greater that the difference among groups is, the better or more distinct the clustering is.

The goal of clustering $L$ points $\{p_1, p_2, \dots, p_L\}$ associated with positive weights $w_1,
w_2,\dots, w_L$ and pairwise distances $d(p_i, p_j)$ into $K$ clusters is to find the $K$ centroids
$\{q_1, q_2,\dots, q_K\}$ of the clusters such that each cluster contains all points that are closer to their centroid than to other centroids:
\begin{equation}
C_i = \{p_l: d(p_l,q_i) \leq d(p_l,q_j), i \neq j\}.
\end{equation}
One can define a clustering cost as a weighted sum of the distances between each point and their corresponding centroid:
\begin{equation}
J = \sum_{i=1}^{K} \sum_{p_j \in C_i} w_j d(q_i, p_j).
\label{equation:generalobject}
\end{equation}
Different choices of pairwise distances $d(p_i, p_j)$ can result in different procedures to obtain
the centroid assignments that minimize the cost $J$~\cite{tan2005introduction}. If the squared
Euclidean distance is used, i.e., $d(p_i, p_j) = \|p_i - p_j \|_2^2$, then each cluster's centroid
will be the weighted mean of its elements, and so the clustering is called $K$-means clustering.  If the
Manhattan distance is used, i.e., $d(p_i, p_j) = \|p_i - p_j \|_1 $, then each cluster's centroid
will be the weighted median of its elements, and so the clustering is called $K$-median clustering.  In the
special case where $p_i \in \mathbb{R}$, i.e., all the points are along a line, and the absolute
value is used as a distance, i.e., $d(p_i,p_j) = |p_i-p_j|$, the cost $J$ is equal to
\begin{equation}
J = \sum_{i=1}^{K} \sum_{p_j \in C_i} w_j |q_i-p_j|.
\label{equation:medianobject}
\end{equation}
Therefore, one can solve for the centroids by setting the subgradient of the measure function
(\ref{equation:medianobject}) with respect to each centroid to zero; the result is
\begin{equation}
\sum_{p_j \in C_i} w_j \text{sign}(q_i-p_j) = 0,
\label{equation:medianproperty}
\end{equation}
for $i = 1,\ldots,K$, where $\text{sign}(x)$ returns the sign of $x$. Equation
(\ref{equation:medianproperty}) illustrates that the resulting centroids are the weighted medians of the
elements in each cluster, and implies that the points on the two sides of the centroids have
maximally balanced weight. That is, for each $i = 1,\ldots,K$,
\begin{equation}
\begin{split}
\sum_{j:p_j \in C_i, p_j \le q_i} w_j &\ge \sum_{j:p_j \in C_i, p_j > q_i} w_j,\\
\sum_{j:p_j \in C_i, p_j < q_i} w_j &\le \sum_{j:p_j \in C_i, p_j \ge q_i} w_j.\\
\end{split}
\label{equation:balanceweight}
\end{equation}

The main difference between $K$-median clustering and the more predominant $K$-means clustering is
that the former minimizes a sum of pairwise Manhattan distances, while the latter minimizes a sum of
squared Euclidean distances. This difference makes the K-median clustering more robust to noise and
outliers since the mean of a cluster deviates from the center of the cluster when outliers are
present, while the median stays close to the center and is less impacted by the outliers. As shown
in the experiments of Section \ref{section:experiments}, $K$-median clustering performs better than
K-means clustering in parameter estimation.

\subsection{Earth Mover's Distance}

The earth mover's distance (\EMD) between two vectors $(c,\hc)$ relies on the notion of mass
assigned to each entry of the involved vectors, with the mass of each entry being equal to its
magnitude. The goal of \EMD~is to compute the lowest transfer of mass among the entries of the first
vector needed in order to match the entries of the second vector. We assume that the vectors $c$ and
$\hc$ are two $K$-sparse vectors with supports $S = \{s_1, s_2, \dots, s_K\}$ and $\hS =
\left\{\hs_1, \hs_2, \dots,\hs_K \right\}$, respectively. $\text{EMD} \left(c,\hc \right)$
represents the distance between the two vectors by finding the minimum sum of mass flows $f_{i, j}$
from entry $c _{s _i}$ to entry $\hc_{ \hs _j }$ multiplied by the distance $d_{i, j} = |s_i-\hs_j|$
that can be applied to the first vector $c$ to yield the second vector $\hc$.\footnote{The standard
definition of \EMD~assumes that the two vectors have the same $\ell_1$ norm. Nonetheless, one can
add additional entries on both vectors to account for the norm mismatch~\cite{pele2009fast}.  More specifically, we denote these additional
entries as $c_{N+1}$ and $\hc_{N+1}$, i.e., both are at the end of the vectors. Their magnitudes are
$c _{N+1} = c _{\min} + \max \left\{ 0, \left\| \hc \right\| _1 - \left\| c \right\| _1 \right\}$ and $\hc
_{N+1} = c _{\min} + \max \left\{ 0, \left\| c \right\| _1 - \left\| \hc \right\| _1 \right\}$,
where $c _{\min}$ is the minimum magnitude among all nonzero entries of $c$ and $\hc$, and the flow
distance between the new entries is $d _{N+1, N+1} = 0$. Furthermore, the flow
distance between all other entries and the new ones are $d _{i, N+1} = d _{N+1, j} = N$ for
both vectors so that the propagation of mass to these new nodes is discouraged and limited to the
mismatch; note that the flow between the two new entries is $f_{N+1,N+1} = c_{\min}$.} 
This is a typical linear programming
problem that can be written as
\begin{equation}
\begin{aligned}
\text{EMD} \left(c,\hc\right) &=&& \min_{f} && \sum_{i,j} f_{i, j}d_{i, j} \\
&&&\text{such that} & &\sum_{j=1}^K f_{i, j} = \left| c_i \right|, \quad i = 1, 2, \dots, K,\\
		 &&&&&\sum_{i=1}^K f_{i, j} = \left| \hc_{j} \right|, \quad j = 1,2,\dots, K,\\
		 &&&&&f_{i, j}  \geq 0, \quad\qquad i,j = 1, 2, \dots K.\\
\end{aligned}
\label{equation:emd}
\end{equation}

\section{Clustering Methods for Parameter Estimation}
\label{section:cpe}

In this section, we present our approach for parameter estimation based on $K$-median clustering.
We begin by studying the relationship between the \EMD~of \PD~coefficient vectors and the error of
parameter estimation in order to justify the use of \EMD~in parameter estimation. Next, we show that
the $K$-median clustering can be used as \EMD-optimal sparse approximation to replace the hard
thresholding operator in standard \CS~recovery algorithms.  Afterward, we provide a theoretical
performance analysis that features an upper bound on the parameter estimation error relying on the
correlation function of the PD elements. Finally, we analyze the effect of compression and noise on
the performance of the proposed algorithm and describe the use of thresholding to mitigate the
performance reduction.

\subsection{Estimation Errors}
\label{section:emd}

We wish to obtain parameter estimates that minimize the total estimation error, i.e., the sum of the
absolute differences $\left| \theta_k - \hp_k \right|$ between each unknown parameter $\theta_k$ and
the corresponding estimated parameter $\hp_k$. Computing the estimation error between a set of $K$
one-dimensional true parameters $\theta = \{\theta_1, \theta_2, \dots, \theta_K\} \subset
\mathbb{R}$ and a set of $K$ one-dimensional estimates $\hp =\left\{ \hp_1, \hp_2, \dots, \hp_K
\right\} \subset \mathbb{R}$ involves an assignment problem that minimizes the cost of assigning
each true parameter to a parameter estimate.  The resulting \PEE~can be obtained by solving the
following linear program, which is a relaxation of the formal integer program optimization
\cite[Corollary~19.2a]{schrijver1998theory}:
\begin{equation}
\begin{aligned}
\text{PEE}(\theta,\hp) &=&& \min_{g} &&\sum_{i,j} g_{i, j}t_{i, j} \\
&&&\text{such that} & &\sum_{j=1}^K g_{i, j} = 1, \quad i = 1, 2, \dots, K;\\
		 &&&&&\sum_{i=1}^K g_{i, j} = 1, \quad j = 1,2,\dots, K;\\
		 &&&&&g_{i, j} \ge 0, \quad\qquad i,j = 1, 2, \dots, K,
\end{aligned}
\label{equation:pee}
\end{equation}
where $t_{i, j} = \left| \theta_i - \hp_j \right|$. The minimizer $g$ for the optimization above
encodes the matching (between the true parameter values and their estimates) that minimizes the total
parameter estimation error,

When the sampling interval of the parameter space that generates by PD is constant and equal to
$\Delta$ (so that $\bar{\theta}_i = \Delta\cdot (i - 1) + \theta_1$, where $i$ denotes the index for
the corresponding PD element $\psi(\bar{\theta}_i)$), it is easy to see that for a pair of true and
estimated coefficient vectors and the corresponding true and estimated parameter values, the computations
of the \PEE~(\ref{equation:pee}) and the \EMD~(\ref{equation:emd}) are a match under the
transformation $t_{ij} = \Delta\cdot d_{ij}$. This straightforward comparison demonstrates the close
relationship between the \EMD~and \PEE~(PEE), which is formally stated in the following theorem and
proven in~\ref{chapter:isometryproof}.

\begin{theorem}
\label{theorem:isometric}
Assume that $\Delta$ is the sampling interval of the parameter space that generates the \PD~used for
parameter estimation. If $c$ and $\hc$ are two $K$-sparse \PD~coefficient vectors with supports
corresponding to two sets of parameters $\theta$ and $\hp$, then the \EMD~between the two
coefficient vectors provides an upper bound to the \PEE~between the two sets of parameters:
\begin{equation}
\text{PEE}(\theta,\hp) \leq \frac{\Delta}{c_{\min}} \text{EMD}(c,\hc),
\label{eq:isometric}
\end{equation}
where $c_{\min}$ is the smallest component magnitude among the nonzero entries of $c$ and $\hc$.
\end{theorem}\par

Theorem~\ref{theorem:isometric} clearly shows that the \EMD~between sparse \PD~coefficient vectors
provides an upper bound of the \PEE~for the corresponding parameters, with a scaling factor
proportional to the \PD~parameter sampling rate $\Delta$. For a particular estimation problem where
the \PD~sampling interval is fixed, Theorem~\ref{theorem:isometric} gives the intuition that, though
minimizing the \EMD~of \PD~coefficient vectors doesn't always minimize the parameter estimation
error, it does minimize its upper bound in (\ref{eq:isometric}) and is likely to return a smaller
estimation error than non-\EMD~based approaches, for which no parameter estimation performance
guarantee exists. Thus, finding an \EMD-optimal sparse approximation algorithm is the next important
step for the proposed parameter estimation method.

It is worth noting from the proof of Theorem~\ref{theorem:isometric} that the relationship between
\EMD~and \PEE~is met with equality when $c_{\min} = c_{\max}$, i.e., all component magnitudes have
the same values. If the dynamic range of component magnitudes is defined as
\begin{equation}
r = \max _{i \ne j} \frac { \left| c_i \right|}{ \left| c_j \right|},
\label{equation:dynamic}
\end{equation}
then $r$ reflects how close to equality (\ref{eq:isometric}) can be. The dynamic range of component
magnitudes is an important condition in parameter estimation, as we will show in the sequel.

\subsection{\EMD-Optimal Sparse Approximation}

As we have just shown, \EMD-optimal sparse approximation plays a crucial role in our proposed
\CPE~approach to enable small estimation error. In a similar way that the hard thresholding operator
return the best sparse approximation vector in terms of Euclidean distance, we present an
\EMD-optimal sparse approximation that will return a sparse approximation vector $\hc \in \bc$ that
is optimal (among all sparse vectors) in terms of the $\text{EMD}\left(\hc, v\right)$ for any input
vector $v \in \bc$. This approach makes it possible to integrate \EMD~into a \CS~framework and to
formulate a new \CPE~algorithm by replacing the hard thresholding operation by the new \EMD-optimal
sparse approximation.

Assume that $I = \{1,2,\dots,L\}$ and $S = \{s_1, s_2,\dots, s_K\}$ is a $K$-element subset of $I$.
Consider the problem of finding the $K$-sparse vector $\hc$ with fixed support $S$ that has the
smallest \EMD~to an arbitrary vector $v$. The minimum flow work defined in the \EMD~is achieved if
and only if the flow is active between each entry of the vector $v$ and its nearest (in terms of index) nonzero entry
$s_i$ of the vector $\hc$. In other words, the nonzero entries of $\hc$ partition the entries of $v$
into $K$ different groups. Denote by $V_i$ the set of the entry indices of $v$ that are matched to
the nonzero entry $s_i$ of $\hc$; this set can be written as
\begin{equation}
V_i = \{l\in I:|l-s_i| \leq |l-s_j|, s_j \ne s_i, s_j \in S\}.
\label{eq:sets}
\end{equation}
The \EMD~defined in (\ref{equation:emd}) can then be written as
\begin{equation}
\text{EMD}(v,\hc) = \sum_{i=1}^K \sum_{j \in V_i} | v_j | | j-s_i |.
\label{equation:emdsparse}
\end{equation}
It is important to note that (\ref{equation:emdsparse}) has the same formula as
(\ref{equation:medianobject}), which is the objective function used in $K$-median clustering. Thus,
one can pose a $K$-median clustering problem to minimize the value of (\ref{equation:emdsparse})
over all possible supports $S$. To that end, define $L$ points in a one-dimensional space with
weights $|v_1|, |v_2|,\ldots, |v_L|$ and locations $1, 2, \ldots, L$, respectively. It is easy to
see that if we denote the set of centroid positions obtained by performing $K$-median clustering for
this problem as $S$, then the set $S$ corresponds to the support of the $K$-sparse signal that is
closest to the vector $v$ when measured with the \EMD. One can then simply compute the sets in
(\ref{eq:sets}) and define the \EMD-optimal $K$-sparse approximation $\hc$ to the vector $v$ as
$\hc_{s_i} = \sum_{j \in V_i} v_j$ for $s_i \in S$, with all other entries equal to zero.

\subsection{Correlation Function in \PD-based Parameter Estimation}

We follow the convention of greedy algorithms for sparse approximation and  \CS, where a proxy of
the coefficient vector is obtained via the correlation of the observations with the \PD~elements,
i.e., $v = \Psi^H x$, where $\Psi ^H$ denotes the Hermitian (conjugate transpose) of $\Psi$. The
resulting proxy vector $v$ can be expressed as a linear combination of shifted correlation
functions. The magnitudes of the components of this linear combination will be proportional to the
magnitudes of the corresponding entries of the sparse coefficient vector $c$.

The correlation value between the \PD~elements corresponding to parameters $\theta_1$ and
$\theta_2$ is defined as
\begin{equation}
\lambda(\omega) = \lambda(\theta_2-\theta_1) = \left| \langle\psi(\theta_2), \psi(\theta_1) \rangle
\right| = \left| \psi^H(\theta_1) \psi(\theta_2) \right|,
\label{equation_correlation}
\end{equation}
where $\omega = \theta_1 - \theta_2$ measures the difference between parameters, and we assume for
simplicity that the correlation depends only on the parameter difference $\omega$. In many parameter
estimation problems, such as frequency estimation and time delay estimation, the correlation
function has bounded variation such that the cumulative correlation function, defined as
$\Lambda(\theta) = \sum_{\omega \in \Thetabar : \omega \le \theta} \lambda(\omega) $,
is bounded. The cumulative correlation function is a nondecreasing function with infimum
$\Lambda(-\infty) = 0$ and supremum $\Lambda(\infty) = \sum_{\omega \in \Thetabar}
\lambda(\omega)$.

\begin{figure}[t]
\centering
\includegraphics[width = 0.4\textwidth]{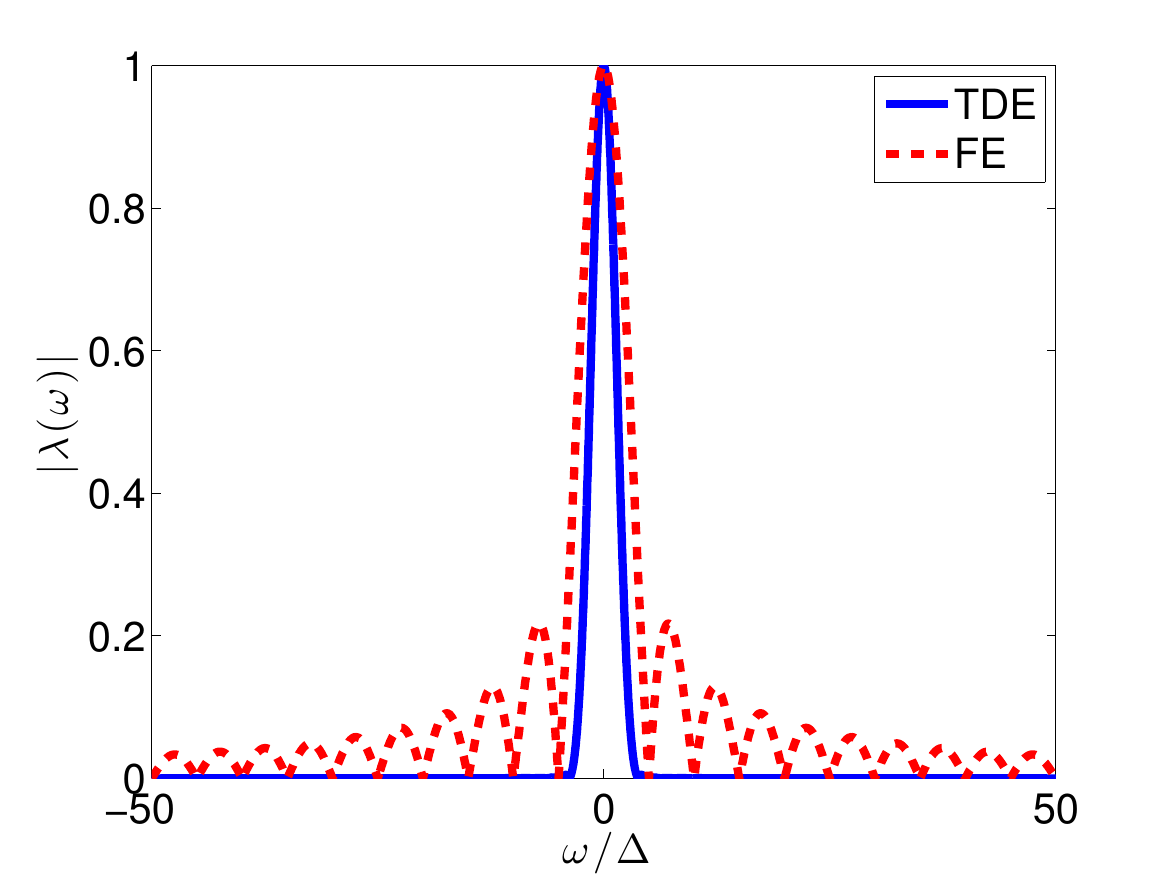}
\includegraphics[width = 0.4\textwidth]{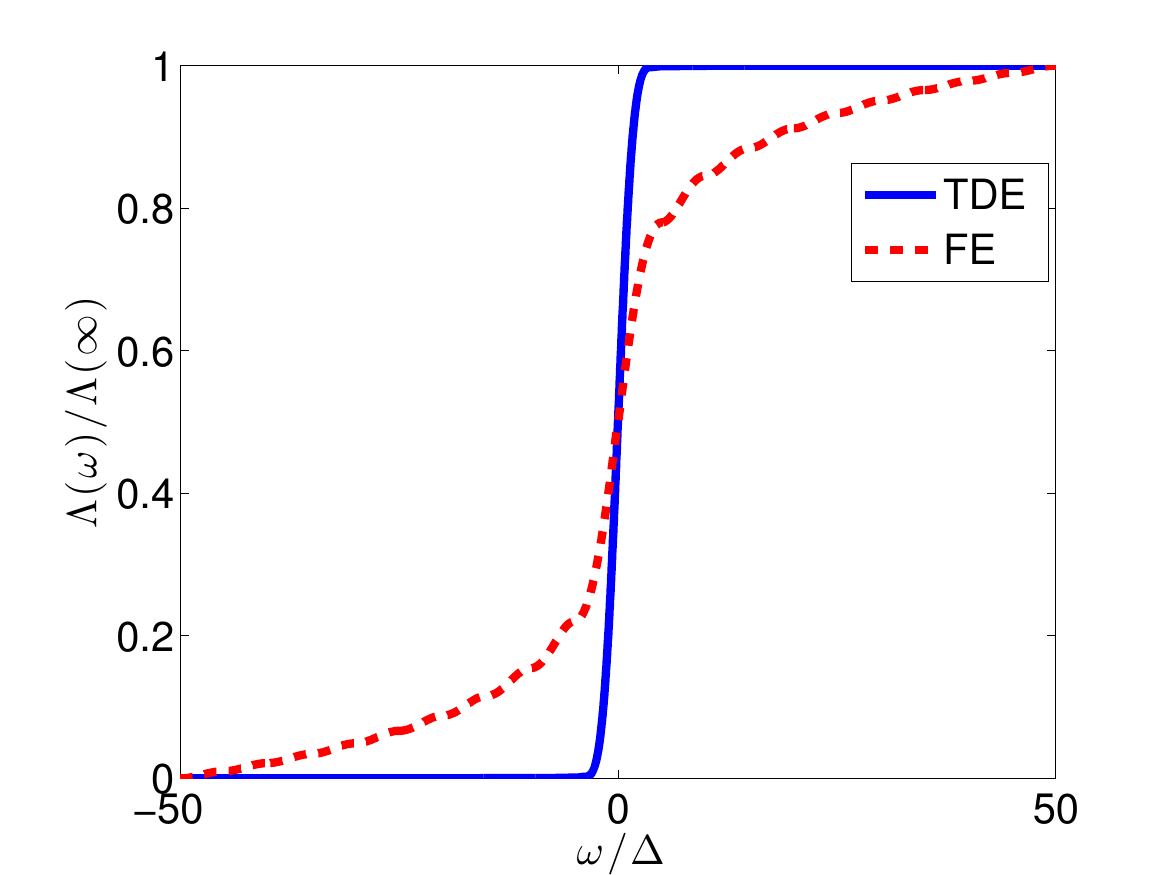}
\caption{\small \sl Examples of correlation in PD constructions. (Left) Correlation function
$|\lambda(\omega)|$ and (right) normalized cumulative correlation function
$\Lambda(\omega)/\Lambda(\infty)$ as a function of the discretized parameter $\omega/\Delta$ for
time delay estimation (TDE) and frequency estimation (FE).}
\label{figure_corr}
\vspace{-5mm}\end{figure}

As shown in Figure~\ref{figure_corr}, the correlation function $\lambda(\omega)$ achieves its
maximum when the parameter difference $\omega = 0$  and decreases as $|\omega|$ increases, finally
vanishing when $|\omega| \to \infty$. In words, the larger the parameter difference is, the smaller
the similarity of the corresponding \PD~elements is. Due to the even nature of the correlation
function, i.e., $\lambda(\omega) = \lambda(-\omega)$, the cumulative correlation function is
rotationally symmetric, i.e., $\Lambda(\theta) + \Lambda(-\theta) = 2 \Lambda(0) = \Lambda(\infty)$,
as shown in Figure~\ref{figure_corr}. Both figures also indicate that the correlation function for
time delay estimation decays much faster than that for frequency estimation, which is indicative of
the increased difficulty for frequency estimation with respect to time delay estimation, as shown in
the sequel.

For convenience of analysis, we assume that the correlation function is real and nonnegative, while
noting that the experimental results match our theory even when the assumption does not hold. When
the signal is measured directly without \CS~compression (i.e., the measurement matrix is the
identity or $\Phi = I$) in a noiseless setting, the observations exactly match the sparse signal and
can be written as
\begin{equation}
y = x = \sum_{i = 1}^K c_i \psi \left( \theta_i \right).
\label{equation_signal}
\end{equation}
Therefore, the proxy entries $v[j] = \psi \left( \thetabar_j \right)^Hx$ correspond to inner
products between the observation vector $y$ and the \PD~elements $\psi \left( \thetabar_j \right)$
corresponding to the sampled parameters $\thetabar_j \in \Thetabar$. Thus, the proxy vector can be expressed as a
linear combination of shifted correlation functions. For $j = 1, 2, \cdots, L$, the $j^\mathrm{th}$ entry of
the proxy vector is equal to
\begin{align}
    v[j] = \psi(\thetabar_j)^Hx =  \sum_{i = 1}^K c_i \psi \left( \thetabar_j \right) ^H \psi
    \left( \theta_i \right) = \sum_{i = 1}^K c_i \lambda \left( \thetabar_j - \theta_i \right).
\label{equation_proxy}
\end{align}
It is easy to see that the proxy vector will feature local maxima at the indices for the sampled
parameters in the \PD~that are closest to the true parameter. Thus, the goal of the parameter
estimation finally reduces to the search of local maxima in the proxy vector over all parameter
samples represented in the \PD, which often is addressed via the sparse approximation operator on
the proxy, e.g., via the hard thresholding operator or \EMD-optimal sparse approximation.

When the correlation function has fast decay (usually the case when the coherence of the \PD~is very
small), it is possible to find the local maxima of the proxy via the hard thresholding operator, as
used in standard greedy algorithms for sparse signal recovery. However, when the correlation
function decays slowly (usually the case when the \PD~elements are highly coherent), the
thresholding operator will unavoidably focus its search around the peak of the proxy with the
largest magnitude, unless additional approaches like band exclusion are implemented. In contrast,
\EMD-optimal sparse approximation identifies the local maxima of the proxy directly by exploiting
the fact that these local maxima correspond to the $K$-median clustering centroids, when certain
conditions (to be defined in Theorem \ref{theorem:generalclustering} in the next subsection) are
met. Thus, we can propose an \EMD-optimal sparse approximation algorithm, shown as Algorithm
\ref{algorithm:clustering}, which leverages a standard iterative, Lloyd-style $K$-median clustering
algorithm \cite{bradley1997clustering} and obtains the optimal sparse approximation of the proxy in
the \EMD~sense. This algorithm will repeatedly assign each entry of the proxy vector to the cluster
whose median is closest to the entry and then update each median by finding the weighted median of
the cluster using the balance property (\ref{equation:balanceweight}). At the end, the parameter
estimates corresponding to the centroids are returned.

\begin{algorithm}[t]
\renewcommand{\algorithmicrequire}{\textbf{Input:}}
\renewcommand{\algorithmicensure}{\textbf{Output:}}
\caption{\small \sl \EMD-Optimal Sparse Parameter Estimator $\hp = \mathbb{C}(v, \Thetabar, K)$}
\label{algorithm:clustering}
\begin{algorithmic}[1]
\REQUIRE PD proxy vector $v$, set of parameter samples $\Thetabar$, target sparsity $K$
\ENSURE parameter estimates $\hp$, sampled indices $S$
\STATE \textbf{Initialize}: set $L = |\Thetabar|$, choose $S$ as a random
$K$-element subset of $\{1,\ldots,L\}$
\REPEAT
\STATE $g_i = \arg\min_{j= 1, \dots, K} | i-s_j |$ for each $i=1,\ldots,L$ \hfill \COMMENT{label
parameter samples}
\STATE $s_j = \operatorname{median} \lbrace \left(i, | v_i | \right) :g_i=j \rbrace$ for each
$j=1,\ldots,K$ \hfill \COMMENT{update cluster medians}
\STATE $\hp = \Thetabar_S$ \hfill \COMMENT{find parameter estimates}
\UNTIL{$S$ does not change or maximum number of iterations is reached}
\end{algorithmic}
\end{algorithm}

\subsection{Performance Analysis}

There are some conditions that the signal $x$ should satisfy to minimize the estimation error when
using Algorithm \ref{algorithm:clustering}.
\begin{itemize}
\item {\em Minimum Parameter Separation:} If two parameters $\theta_i$ and $\theta_j$ are too close
    to each other, the similarity of $\psi(\theta_i)$ and $\psi(\theta_j)$ makes it difficult to
    distinguish them. Therefore, our first condition considers the minimum separation distance:
\begin{equation}
\zeta = \min_{i \neq j} |\theta_i-\theta_j|.
\label{equation:separation}
\end{equation}
\item {\em Parameter Range:} It is often convenient to restrict the feasible parameters and sampled
    parameters in a small range, i.e, $\Thetabar$ is bounded. Any parameter observed should be
    sufficiently far away from the bounds of the parameter range. According to
    (\ref{equation:balanceweight}), which implies a balance of the proxy $v$ around the local maxima
    when using a clustering method, there will be a bias in the estimation due to the missed portion
    of the symmetric correlation function $\lambda$ when the unknown parameter is too close to the
    bound of the parameter range.  Therefore, the condition should also consider the minimum
    off-bound distance $\epsilon$, formally written as
\begin{equation}
\epsilon = \min_{1\le i \le K} \left\{ \min \left( \theta_i - \min \left( \Thetabar \right), \max
    \left( \Thetabar \right) - \theta_i \right) \right\}.
\label{equation:offbound}
\end{equation}
\item {\em Dynamic Range:} If the magnitudes of some components in the signal are too small, they
    may be dwarfed by larger components and ignored by the greedy algorithms. Thus, we need to pose
    an additional condition on the dynamic range of the component magnitudes as defined in
    (\ref{equation:dynamic}).
\end{itemize}

With these conditions, we can formulate the following theorem that guarantees the performance of
our clustering-based method for parameter estimation.
\begin{theorem}
\label{theorem:generalclustering}
Assume that the sampling interval of the parameter space $\Delta \rightarrow 0$, and that the signal $x$
given in (\ref{equation_signal}) involving $K$ parameters $\theta_1,
\theta_2, \dots, \theta_K$ has a dynamic range of the component magnitudes as defined in
(\ref{equation:dynamic}). For any allowed
error $\sigma > 0$, if the minimum separation distance (\ref{equation:separation}) satisfies
\begin{equation}
\zeta \geq 2\Lambda^{-1}\left(2 \Lambda(0) \left(1 - \frac{\Lambda(\sigma) / \Lambda(0) - 1}{( 2 K -
2) r + 1}\right)\right) + 2\sigma,
\label{equation:miniSep}
\end{equation}
and the minimum off-bound distance (\ref{equation:offbound}) satisfies
\begin{equation}
\epsilon \geq \Lambda^{-1}\left(2 \Lambda(0) \left(1 - \frac{\Lambda(\sigma) / \Lambda(0) -
1}{2Kr}\right)\right),
\end{equation}
then the estimates $\hp_1, \hp_2, \dots, \hp_K$ returned from Algorithm \ref{algorithm:clustering}
have estimation error
$\text{PEE} \left( \theta, \hp \right) \leq K \sigma$.
\end{theorem}

The full proof of Theorem~\ref{theorem:generalclustering} is provided in \ref{chapter:clusteringproof}.
Theorem \ref{theorem:generalclustering} provides a guarantee on the parameter estimation error
obtained from $K$-median clustering of the proxy $v$ for the \PD~coefficients. Theorem
\ref{theorem:generalclustering} also provides a connection between the conditions described earlier
and the performance of \PD-based parameter estimation, providing a guide for the design and
evaluation of \PDs~that can achieve the required estimation error for practical problems. For
example, in time delay estimation problems, instead of increasing the minimum separation distance
required for recovery, one can try to design a transmitted waveform that improves the other
conditions cited above (such as one that increases the rate of decay of the cumulative correlation
functions, as will be discussed in the sequel) to achieve small estimation error.

Theorem \ref{theorem:generalclustering} also makes explicit the linear relationship between
the normalized cumulative correlation for the minimum separation distance $\Lambda(\zeta /
2)/\Lambda(\infty)$ and the normalized cumulative correlation for the maximum observed error
$\Lambda(\sigma)/ \Lambda(0)$, which can be obtained by applying $\Lambda(\cdot)$ on both sides of
(\ref{equation:miniSep}) while ignoring the small term $\sigma$. Additionally, the required minimum
separation distance will be dependent on the specific parameter estimation problem, even when the
maximum allowed error is kept constant. This illustrates the wide difference in performances 
in parameter estimation, e.g., between time delay estimation and frequency estimation: 
the minimum separation distance required by time delay estimation is much smaller than that 
of frequency estimation due to the contrasting rates of decay of the function $\Lambda$, as 
shown in Figure~\ref{figure_corr}. Once more, we emphasize that
although Theorem \ref{theorem:generalclustering} is derived for the case of a nonnegative real
correlation function and is asymptotic on the parameter sampling interval $\Delta$, our numerical
simulations in the sequel show that the predicted relationship between $\zeta$ and $\sigma$ are
observed in practical problems of modest sizes.

\subsection{Effect of Compression}

The addition of \CS~and measurement noise make the estimation problem more difficult, since in both
of these cases there is a decrease in the rate of decay of the cumulative correlation function
$\Lambda$, as shown in the sequel. When the measurement matrix $\Phi$ is used to obtain the observed
measurements $y$ from the signal $x$ such that $y = \Phi x = \sum_{i=1}^K c_i \Phi \psi(\theta_i)$,
the proxy becomes
\begin{align}
    v[j] = \langle \Phi^H y, \psi(\thetabar_j) \rangle = \langle y, \Phi \psi(\thetabar_j)
    \rangle = \sum_{i=1}^K c_i \langle \Phi \psi(\theta_i) , \Phi \psi (\thetabar_j) \rangle.
\label{equation_phi_proxy}
\end{align}
Only if $\Phi^H\Phi = I$ can (\ref{equation_phi_proxy}) be identical to (\ref{equation_proxy}). We
define the compressed correlation function as
\begin{align}
\lambda_{\Phi} \left( \omega \right) &= \lambda_{\Phi} (\theta_1 - \theta_2) = \langle \Phi
\psi(\theta_2) , \Phi \psi(\theta_1) \rangle = \psi^H(\theta_1) \Phi^H\Phi \psi(\theta_2).
\label{equation_induced_correlation}
\end{align}
The proxy can again be expressed as the linear combination of shifted copies of the redefined
correlation function (\ref{equation_induced_correlation}):
\begin{equation}
v[j] = v\left( \thetabar_j \right) = \sum_{i=1}^K c_i \lambda_{\Phi} \left( \thetabar_j - \theta_i \right).
\label{equation_induced_proxy}
\end{equation}

Although in general we will have $\lambda \ne \lambda_\Phi$, we can use the preservation property of
inner products through random projections~\cite{vempala2004the-random}. That is, when $\Phi$ has
independent and identically distributed (i.i.d.) Gaussian random entries and sufficiently many rows,
there exists a constant $\delta > 0$ such that for all pairs $(\theta_i,\theta_j)$ of interest we
have
\begin{equation}
(1-\delta) \lambda (\theta_i - \theta_j) \le \lambda_{\Phi} (\theta_i - \theta_j) \le (1+\delta) \lambda (\theta_i - \theta_j).
\label{eq:ipp}
\end{equation}
The parameter $\delta$ decays as the compression rate $\kappa = M/N$ increases, and the manifolds
$\lambda(\cdot)$ we consider here are known to be amenable to large amounts of compression
\cite{eftekhari2014new-analysis}. Such a relationship indicates that the compression can affect the
correlation function and, by extension, the performance of clustering methods for \CPE. 

\subsection{Quantifying the Role of Correlation Decay}

We choose to focus on simple bounds for the correlation function $\lambda$ to analyze its role in
the performance of \EMD~and \PD-based parameter estimation. Similarly to~\cite{tang2013atomic}, we
use bounding functions to measure and control the decay rate of the correlation function $\lambda$.
We approximate the correlation function $\lambda_\Phi(\omega)$ with an exponential function
$\bar{\lambda}(\omega) = \exp\left(-a \left| \omega \right| \right)$ that provides an upper bound of
the actual correlation function, i.e., $\lambda_\Phi(\omega) \le \bar{\lambda}(\omega)$. The
performance obtained from the exponential function approximation $\bar{\lambda}(\cdot)$ provides an
upper bound of the performance from the real correlation function $\lambda(\cdot)$. In the
exponential function, $a$ is the parameter that controls the decay rate: the larger $a$ is, the
faster that the correlation function decays.  The decay rate of the compressed correlation function
(\ref{equation_induced_correlation}) will be smaller than that of the original correlation function
(\ref{equation_correlation}). We assume that $\lambda \left( \omega \right) = \exp\left(-a \left|
\omega \right| \right)$ and that a bound $\lambda _\Phi \left( \omega \right) \le \exp\left(- b
\left| \omega \right| \right)$ exists; in this case, $b < a$ due to the fact that  (\ref{eq:ipp})
provides us with the following upper bound:
\begin{align}
\lambda _\Phi \left( \omega \right) & \le (1+\delta) \lambda \left( \omega \right) \le (1 +
\delta)\exp\left( -a \left| \omega \right| \right) \le \exp\left( - \left(a - \frac{\ln (1+\delta)}{\left| \omega \right|} \right) \left| \omega
\right| \right),
\end{align}
where $\ln (1+\delta) / \left| \omega \right| > 0$ when $\delta > 0$. This shows that \CS~reduces
the decay speed of the correlation function and increases the necessary minimum separation distance
and minimum off-bound distance to guarantee the preservation of parameter estimation performance.
This dependence is also manifested in the experimental results of Section \ref{section:experiments}
when the correlation function does not follow an exact exponential decay.

We observe in practice that the issues with slow-decaying correlation functions arise whenever the
sum of the copies of the correlation functions far from their peaks becomes comparable to the peak
of any given copy. Thus, one can use operators such as thresholding functions to remove this effect
from appearing in Algorithm \ref{algorithm:clustering}. We can write a hard-thresholded version of
the proxy from (\ref{equation_induced_proxy}) as
\begin{equation}
v_t(\theta_i) = \left\{ \begin{array}{ll}
v(\theta_i) & | v(\theta_i) | > t,\\
0 & v(\theta_i) | \le t,
\end{array} \right.
\label{equation_thresholding}
\end{equation}
where $t$ is the value of the thresholding. The following theorem is proven in
\ref{chapter:performanceproof} and focuses on the minimum separation $\zeta$ by assuming that
$\epsilon$ is sufficiently large. The theorem demonstrates that the thresholding operator reduces
the required minimum separation distance for accurate estimation.
\begin{theorem}
\label{theorem:performance}
Under the setup of Theorem \ref{theorem:generalclustering}, assume that the correlation function
defined in (\ref{equation_induced_correlation}) is given by $\lambda_\Phi \left(\omega\right) =
\exp\left( - a | \omega | \right)$. For any allowed error $\sigma > 0$, when the minimum off-bound
distance $\epsilon$ is large enough, if the dynamic range given in (\ref{equation:dynamic}) is equal
to $r$, the threshold given in (\ref{equation_thresholding}) satisfies $rc_{\min} \exp (-a\sigma)
\le t \le c _{\min}$, and the minimum separation distance given in (\ref{equation:separation})
satisfies
\begin{equation} 
    \zeta \geq \frac{1}{a} \ln \left( \sqrt{ \frac{8 r^2}{ t^2/\left(r c_{\min} \right)^2 -
    \exp(-2a\sigma) } }+1 \right)
\end{equation}
where $c_{\min}$ is the minimum component magnitude, then the estimates $\hp_1, \hp_2, \dots, \hp_K$
returned from performing Algorithm~\ref{algorithm:clustering} on the thresholded proxy $v_t$
in~(\ref{equation_thresholding}) have estimation error
$\text{PEE} \left( \theta, \hp \right) \leq K \sigma$.
\end{theorem}

Theorem \ref{theorem:performance} extends Theorem \ref{theorem:generalclustering} 
by including the use of thresholding as a tool to combat the slow decay of the correlation 
function, due to an ill-posed estimation problem or the use of \CS. One can also
instinctively see that the presence of noise in the measurements will also slow the decay of
the correlation function, which according to the theorem will require larger minimum 
separation or careful thresholding. In practice, the decay coefficient $a$ can usually be 
obtained by finding the minimum value such that the exponential function 
$\exp \left( - a | \omega | \right)$ provides a tight upper bound for the correlation function 
$\lambda(\omega)$.

Perhaps the most important application of
Theorem~\ref{theorem:performance} focuses on the choice of threshold $t$ for the particular problem
of interest, which can improve the performance of the clustering method in \CPE. To deal with the
problems of slow decay or large dynamic range, one can try to increase the threshold value on the
proxy rather than increasing the minimum separation to improve the estimation performance. Again,
although Theorem 3 is based on an approximation of the actual \CPE~problem setup, our numerical
results in the sequel show its validation in practical settings for time delay estimation and
frequency estimation.

It is worth noting the similarities between our analysis above and that performed
in~\cite{tang2013atomic}. Both analyses rely on the assumption that the cumulative correlation
function is symmetric, nondecreasing and variation-bounded. Additionally, both analyses use an upper
bound of the correlation function to simplify the analysis. Furthermore, both analyses derive the
requirement that the separation between any two parameters needs to be large enough. The main
difference between the analyses is that the guarantees of~\cite{tang2013atomic} are focused on exact
parameter estimation, while our analysis considers the quality of parameter approximation when the
problem setup does not provide exact estimation.

\section{Numerical Experiments}
\label{section:experiments}

In order to test the performance of the clustering parameter estimation method on different
problems, we present a number of numerical simulations involving time delay estimation and 
frequency estimation.\footnote{Additional experiments are available in an early version of this manuscript~\cite{mo2014performance}.} Before detailing our experimental setups, we define the parametric signals and the
\PDs~involved in these two example applications.

The parametric signal model for time delay estimation uses a chirp signal $g(t) = \exp
\left( j 2 \pi \left( f_c + f_a \frac{t}{T} \right) t \right) p(t)$, where
\begin{equation}
p(t) = \left\{\begin{aligned} 
        & \sqrt{\frac{2}{3T f_s}} \left(1 + \cos \left(2 \pi \frac{t}{T} \right)\right), & t \in [0,
    T],\\ 
    & 0, & \text{otherwise.}
\end{aligned}\right..
\end{equation}
The length of the chirp and the chirp's starting frequency are fixed to be $T=1~\mu\text{s}$ and
$f_c = 1~\text{MHz}$ across all experiments. The chirp's frequency sweep $f_a$ will have different
values in different experiments. The sampling frequency of the chirp
signal is $f_s = \frac{1}{T_s}$, and $N$ samples are taken for each signal. The delay 
parameter space range goes from $\theta_{\min} = 0$ to $\theta_{\max} = NT_s$. 
The \PD~for time delay estimation contains all chirp signals corresponding to the 
parameter space sampled at a resolution of $\Delta$, e.g., 
\begin{align}
\Psi = [\psi(0), \psi(\Delta), \dots, \psi(\theta_{\max})], 
\label{eq:pd}
\end{align}
with entries
\begin{align}
    \psi(\theta)[n] = g(nT_s - \theta), n = 0, 1, \dots, N - 1.
\end{align}

For frequency estimation, the parametric signals are the $N$-dimensional complex exponentials with entries
\begin{equation}
\psi(\theta)[n] = \frac{\exp\left( j 2 \pi \theta \frac{n}{N} \right)}{\sqrt{N}} , \quad n = 0, 1,
\dots, N-1.
\end{equation}
The parameter space range goes from $\theta_{\min} = 0$ to $\theta_{\max} = N~\text{Hz}$. As before,
the \PD~for frequency estimation is described in (\ref{eq:pd}) with sampling interval
$\Delta$. In both time delay estimation and frequency estimation, the number of unknown parameters
is set to $K = 4$.

\subsection{Theorem Validation}

In the first experiment, we illustrate the relationship between the minimum separation distance and
the maximum allowable error described in Theorem \ref{theorem:generalclustering}. We measure the
performance for each minimum separation distance $\zeta$ by the maximum estimation error $\sigma$
over $1000$ signals with randomly chosen parameter values that are spaced by at least $\zeta$. For
time delay estimation, $f_a = 20~\text{MHz}$, $f_s = 50~\text{MHz}$, $N = 500$, and $\Delta =
0.001~\mu\text{s} $ so that the \PD~contains observations for $10001$ parameter samples, and we let
the minimum separation $\zeta \in \left[ 0.07~\mu\text{s}, 0.1~\mu\text{s} \right]$. For frequency
estimation, $N = 500$ and $\Delta = 0.05~\text{Hz}$, so that the \PD~contains observations for 10001
parameter samples, and we let the minimum separation $\zeta \in \left[ 35~\text{Hz}, 75~\text{Hz}
\right]$.

\begin{figure}[t]
    \centering
\includegraphics[width = 0.4\textwidth]{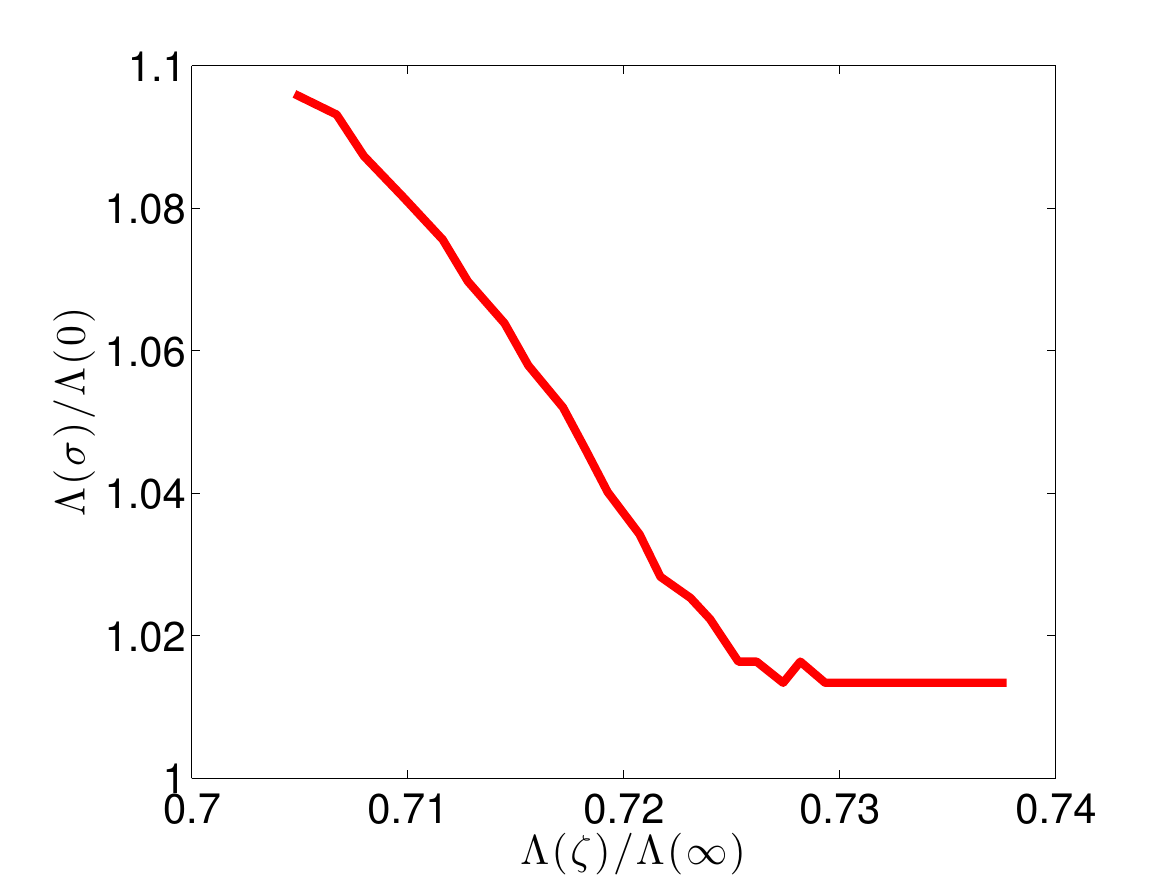}
\includegraphics[width = 0.4\textwidth]{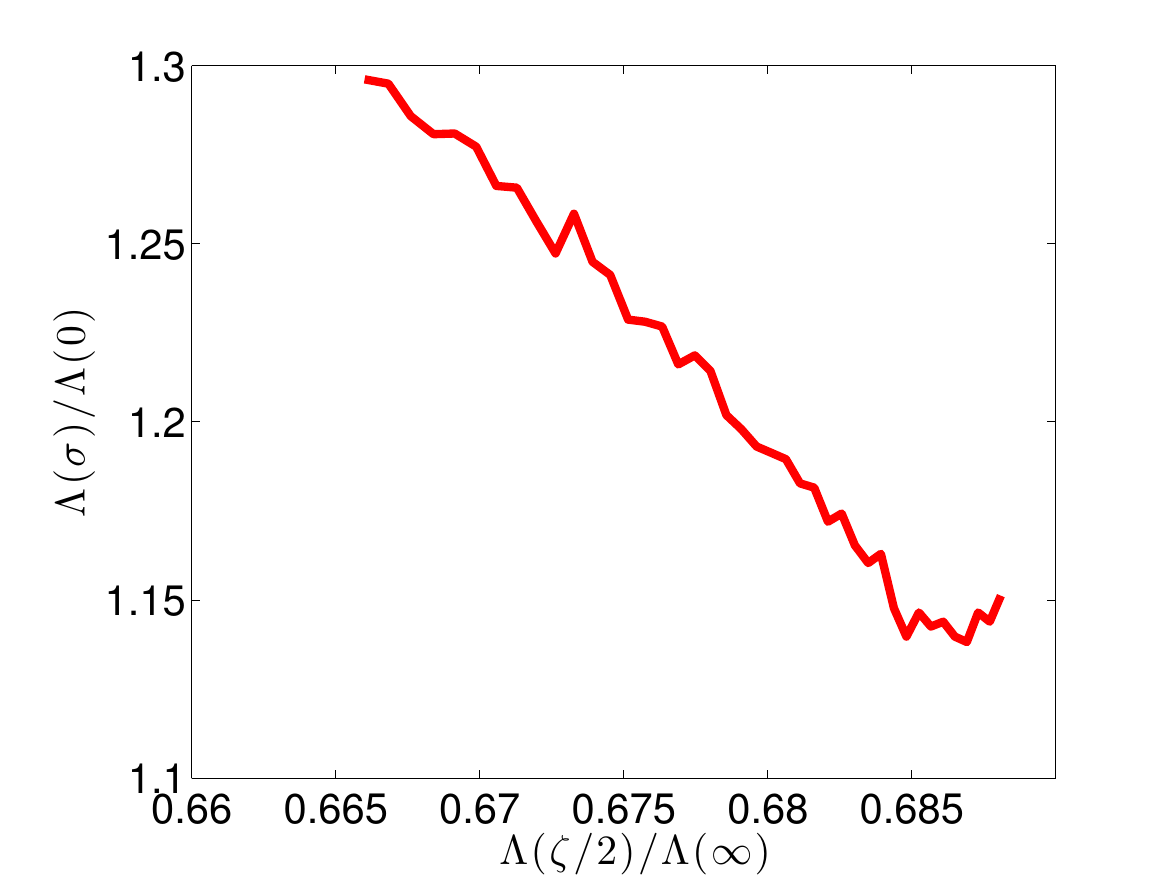}
\includegraphics[width = 0.4\textwidth]{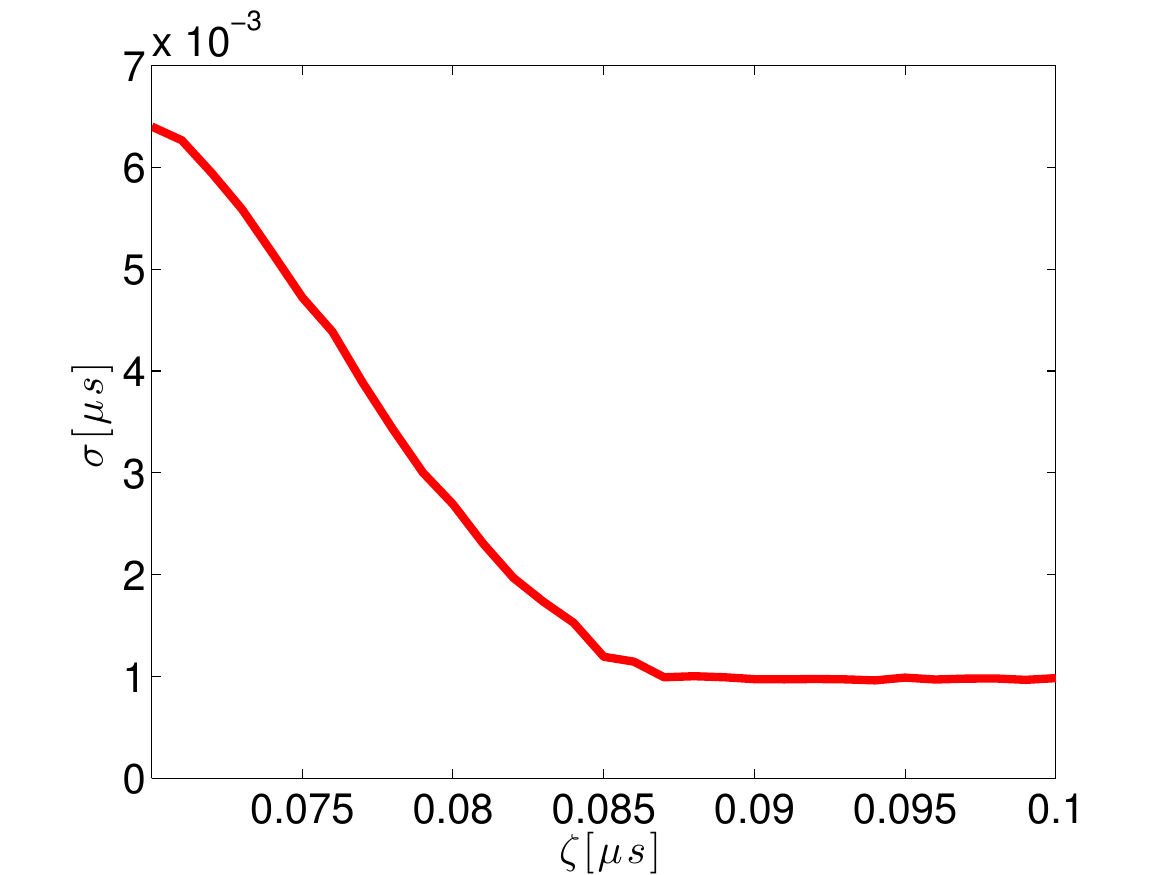}
\includegraphics[width = 0.4\textwidth]{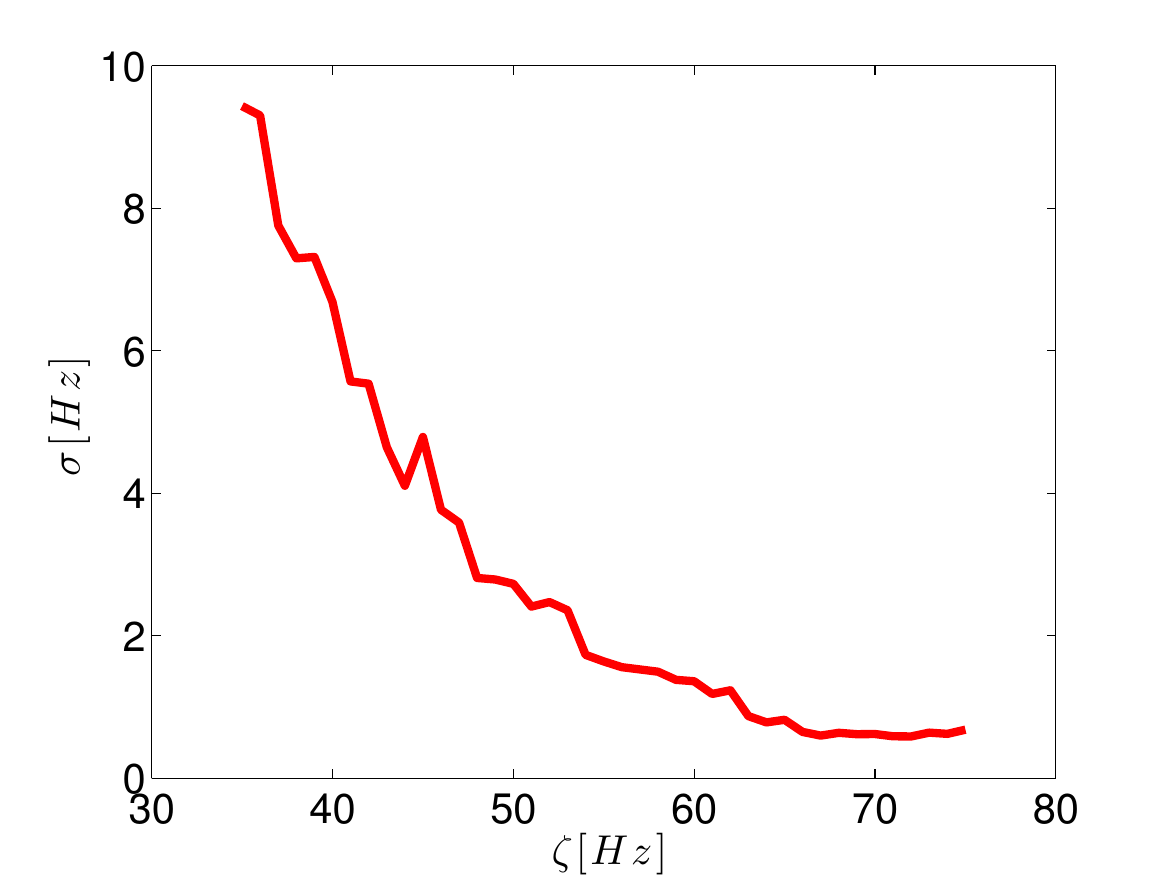}
\caption{\small \sl Performance results for two parameter estimation problems as a function of the
    parameter separation. Left: \TDE; right: \FE. Top: Normalized cumulative correlation function
    for the maximum  estimation error $\Lambda(\sigma)/\Lambda(0)$ as a function of the normalized
    cumulative correlation function for the minimum separation $\Lambda(\zeta / 2)/\Lambda(\infty)$,
    showing a linear dependence. Bottom: Maximum error $\sigma$ as a function of Minimum separation
$\zeta$.}

\label{figure:condition}
\vspace{-5mm}
\end{figure}

Figure~\ref{figure:condition} shows the normalized cumulative correlation for the maximum error
$\Lambda(\sigma)/ \Lambda(0)$ as a function of the normalized cumulative correlation for the minimum
separation distance $\Lambda(\zeta/2)/\Lambda(\infty)$ for both time delay estimation and frequency
estimation. The figure also shows the relationship between the minimum separation $\zeta$ and the
maximum error $\sigma$ without the use of the cumulative correlation function for both example
cases. The approximately linear relationship between $\Lambda(\zeta)/\Lambda(\infty)$ and
$\Lambda(\sigma)/ \Lambda(0)$ for both the time delay estimation and frequency estimation cases
numerically verifies the result of Theorem \ref{theorem:generalclustering}. The difference between
the performance results, as well as the relationship between $\zeta$ and $\sigma$, validates the
conclusion that frequency estimation requires a significantly larger minimum separation than time
delay estimation. We also see from Figure~\ref{figure:condition} that it is impossible to get an
arbitrarily small estimation error even if the minimum separation keeps increasing, as the
estimation error cannot be smaller than the parameter sampling step $\Delta$ (as observed in the
figure). In fact, the figure shows that the relationship between $\Lambda(\zeta/2) /
\Lambda(\infty)$ and $\Lambda(\sigma) / \Lambda(0)$ ends its linearity exactly when the error
approach the parameter sampling resolution $\Delta$ for both application examples. To achieve
precision above the resolution $\Delta$, the use of additional methods such as interpolation is
needed \cite{ekanadham2011recovery}. Nonetheless, though the function $\Lambda(\cdot)$ varies 
in different parameter estimation problems, we consistently obtain a linear relationship between the
minimum separation distance and the parameter estimation error for several parameter
estimation problems by employing the function $\Lambda(\cdot)$.

\begin{figure}[t]
\hspace{-0.6cm}
\includegraphics[width = 0.35\textwidth]{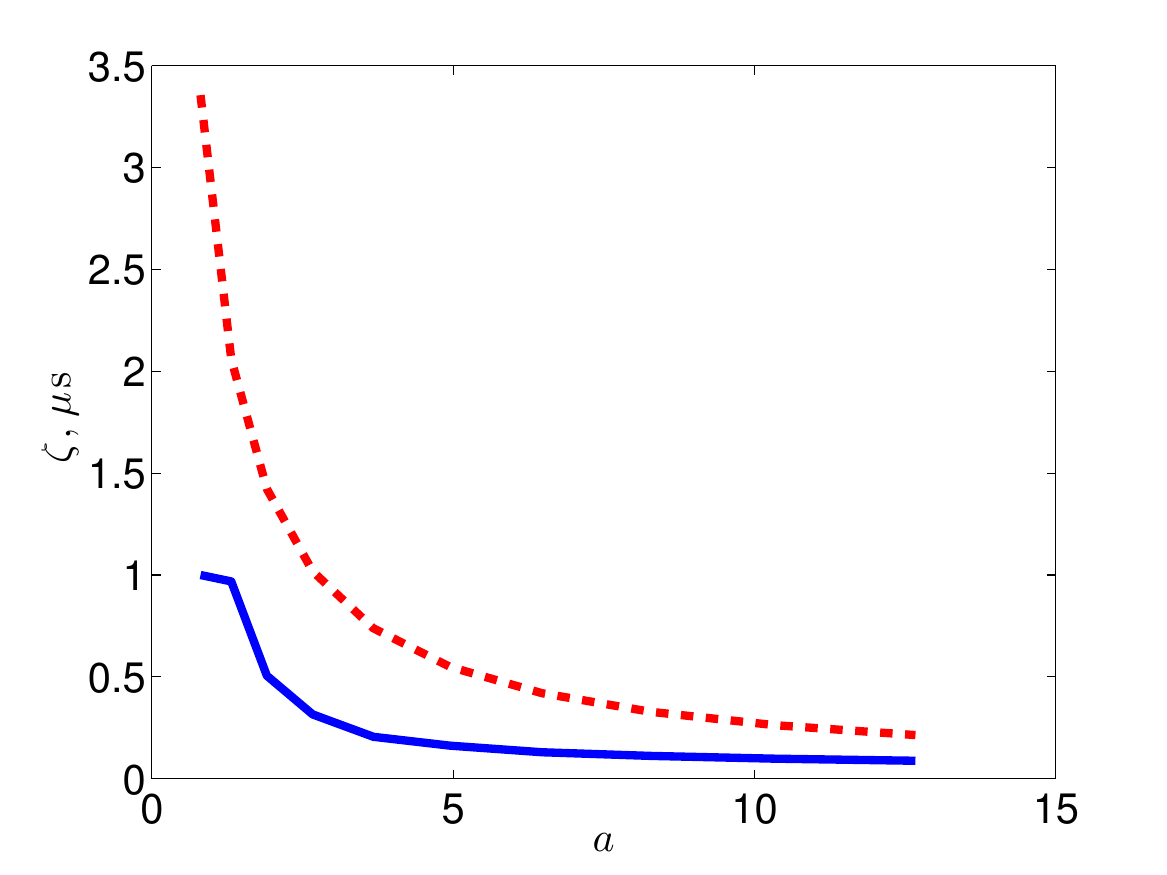}
\includegraphics[width = 0.35\textwidth]{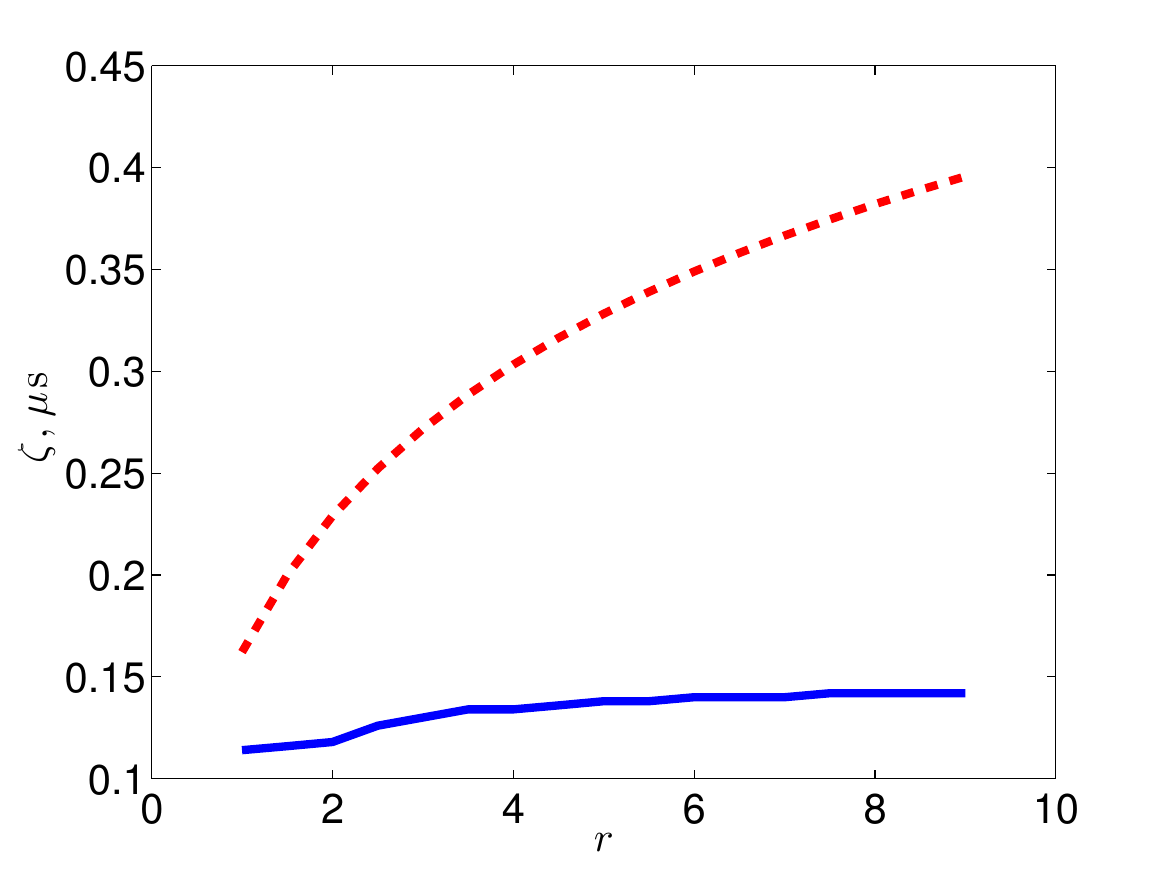}
\includegraphics[width = 0.35\textwidth]{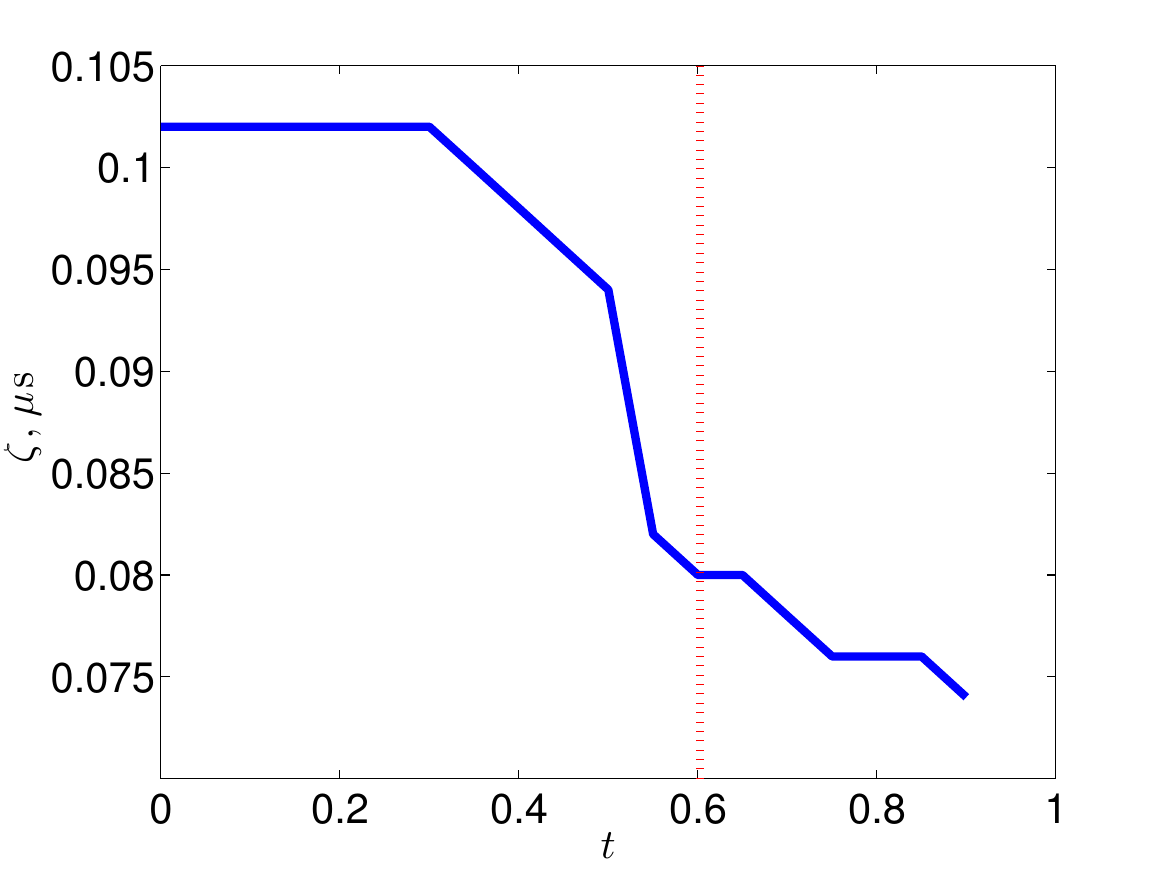}
\caption{\small \sl Performance results for time delay estimation. 
The plots show the minimum separation necessary for accurate estimation as a function of (left) decay coefficient $a$,
(middle) dynamic range $r$, and (right) threshold value $t$. Blue solid curves denote experimental results, while red dashed curves denote the bounds given by Theorem \ref{theorem:performance}.}
\label{figure_TDE_condition}
\vspace{-2mm}
\end{figure}

In the second experiment, we illustrate the application of Theorem \ref{theorem:performance} in the
time delay estimation problem. While the values of all other parameters are kept the same as in the
first experiment, we set $\Delta = 0.002~\mu \text{s}$ and vary the chirp frequency sweep $f_a$
between $2~\text{Hz}$ and $20~\text{Hz}$ to generate different rates of decay of the correlation
function. For each chirp frequency sweep, we obtain the decay parameter $a$ as the smallest value
that enables the exponential function $\exp\left( - a | \omega | \right)$ to bound the correlation
function $\lambda \left( \omega \right)$. We then determine the minimum separation $\zeta$ for which
a maximum estimation error $\sigma = 0.04~\mu\text{s}$ is obtained for
Algorithm~\ref{algorithm:clustering} over 1000 randomly drawn signals having the given minimum
separation $\zeta$. The first result in Figure~\ref{figure_TDE_condition} shows the approximate
reciprocal relationship between $\zeta$ and $a$, which is very close to the line $\zeta = \ln \left(
\sqrt{8} / 0.2 + 1 \right) / a = 2.71/a$ predicted by Theorem \ref{theorem:performance}, when
the threshold level $t = 0.2$ and dynamic range $r = 1$. The second result shows the approximate
logarithmic relationship between the minimum separation $\zeta$ and the dynamic range $r$, when
we set the decay parameter $a = 8.28$ and the threshold $t = 0.9$. The minimum separation observed
is smaller than the prediction $\zeta = \ln \left( \sqrt{8} r ^2 / 0.9 + 1 \right) / 8.28 = \ln
\left( 3.1 r ^2 + 1 \right) / 8.28$ given by Theorem \ref{theorem:performance}, which is likely
explained by the difference between the correlation function and its exponential upper bound. The
third result in Figure~\ref{figure_TDE_condition} shows that required threshold for accuracy
estimation when $a = 12.67$ and $r = 1$ is slightly less than the value $t  = \exp \left( -12.67
\cdot 0.04 \right) = 0.60$ predicted by the theorem. The results presented above can be interpreted
as a numerical corroboration of Theorem~\ref{theorem:performance}.

\subsection{Performance Comparison}

Our following experiments test the performance of clustering methods in \CPE~for different settings.
We incorporate $K$-median clustering into subspace pursuit, a standard sparse recovery algorithm
introduced in \cite{dai2009subspace}, by replacing each instance of the hard thresholding operator
in subspace pursuit with an instance of Algorithm~\ref{algorithm:clustering}. The resulting
clustering subspace pursuit (CSP) is shown in Algorithm \ref{algorithm:estimation}, where
$\mathbb{C}$ refers to Algorithm \ref{algorithm:clustering} and $M^+$ denotes the Moore-Penrose
pseudoinverse of $M$.  Similarly to the band-exclusion subspace pursuit (BSP) used in
\cite{fyhn2013compressive, duarte2013spectral, fyhn2013spectral}, CSP repeatedly computes the proxy
for the coefficient vectors from the measurement residual and then obtains the indices of the
potential parameter estimates from the thresholded proxy using Algorithm \ref{algorithm:clustering}.
A second clustering refines the estimation after the potential estimates are merged with the
previous estimates in order to maintain a final set of $K$ estimates at the end of each iteration.
CSP can also be armed with polar interpolation to significantly improve the estimation precision in
a manner similar to band-excluded interpolating subspace pursuit
(BISP)~\cite{mo2013compressive,ekanadham2011recovery}; we call the resulting algorithm clustering
interpolating subspace pursuit (CISP).
\begin{algorithm}[t]
\renewcommand{\algorithmicrequire}{\textbf{Input:}}
\renewcommand{\algorithmicensure}{\textbf{Output:}}
\caption{\small \sl Clustering Subspace Pursuit (CSP)}
\label{algorithm:estimation}
\begin{algorithmic}[1]
\REQUIRE measurement vector $y$, measurement matrix $\Phi$, sparsity $K$, set of sampled parameters $\Thetabar$, threshold $t$
\ENSURE estimated signal $\widehat{x}$, estimated parameter values $\hp$
\STATE \textbf{Initialize:} $\widehat{x} = 0$, $S = \emptyset$, generate PD $\Psi$ from $\Thetabar$.
\REPEAT
\STATE $y_r = y - \Phi \widehat{x}$ \hfill \COMMENT{Compute residual}
\STATE $v = (\Phi\Psi)^H y_r$ \hfill \COMMENT{Obtain proxy from residual}
\STATE $v( |v| \le t) = 0$ \hfill \COMMENT{Threshold proxy}
\STATE $S = S \cup \mathbb{C}(v, \Thetabar, K)$ \hfill \COMMENT{Augment parameter estimates from proxy}
\STATE $c = (\Phi \Psi_S)^+ y$ \hfill \COMMENT{Obtain proxy on parameter estimates}
\STATE $S = \mathbb{C}(c, \Thetabar_S, K)$ \hfill \COMMENT{Refine parameter estimates}
\STATE $\widehat{x} = \Psi_S c_S$ \hfill \COMMENT{Assemble signal estimate}
\STATE $\hp = \Thetabar_S$ \hfill \COMMENT{Assemble parameter estimates}
\UNTIL{a convergence criterion is met}
\end{algorithmic}
\end{algorithm}

\begin{figure}[t]
\centering
\includegraphics[width = 0.4\textwidth]{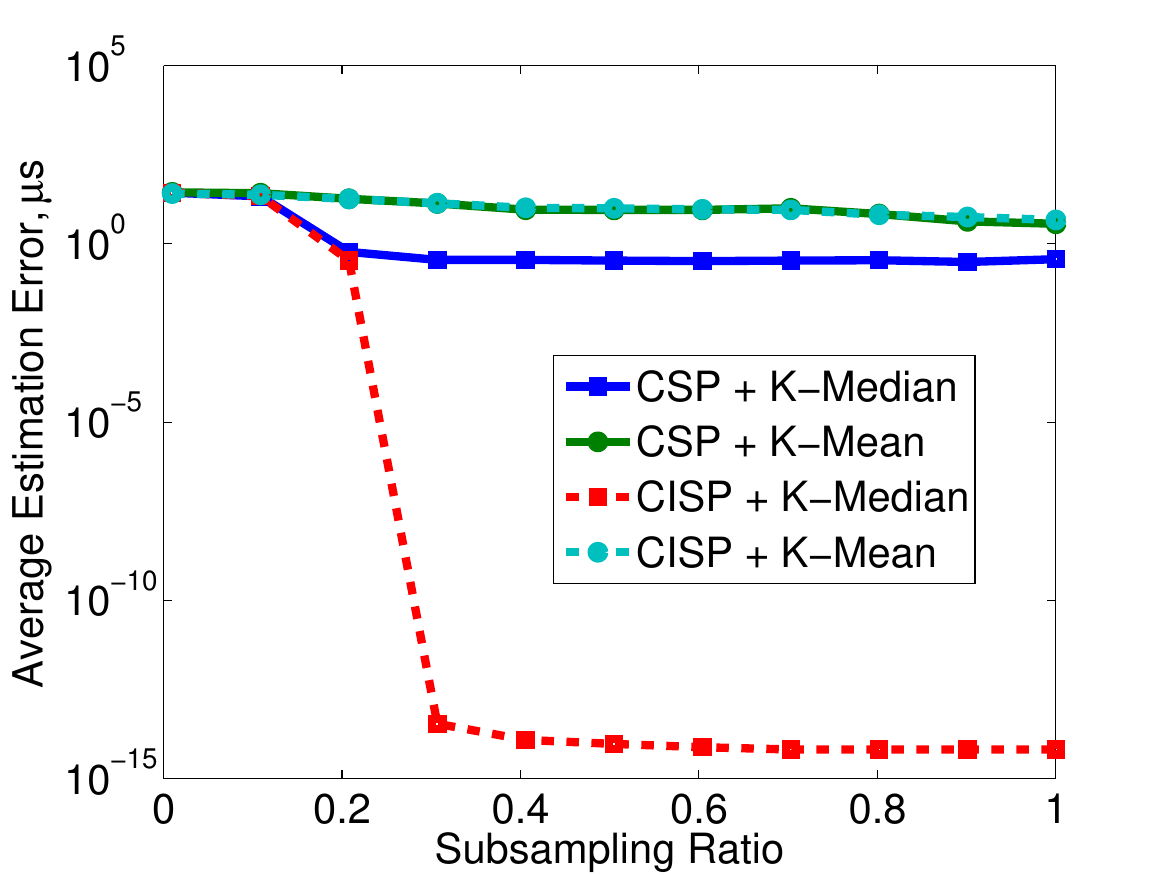}
\caption{\small \sl Average time delay estimation error of Algorithm~\ref{algorithm:estimation} with
$K$-median clustering and $K$-means clustering.}
\label{figure:mean}
\vspace{-2mm}
\end{figure}

Again, in Algorithm~\ref{algorithm:estimation}, we use the $K$-median clustering rather than
$K$-means clustering because the criteria of $K$-median clustering to minimize Manhattan distance
matches the definition of \EMD~and is more robust to the presence of outliers, which can be caused
by a slowly decaying correlation function, compression, or noise.  Figure~\ref{figure:mean} shows
the average time delay estimation error of two versions of Algorithm~\ref{algorithm:estimation}: one
that uses $K$-median clustering and another one that uses $K$-means clustering. It is clear that
$K$-means clustering is not able to obtain the same estimation quality as $K$-median clustering.

Besides BSP and CSP, we also evaluate the performance on approximate message passing with MUSIC
(AMP) \cite{Hamzehei2015Compressive-Par, Hamzehei2016Compressive-Dir} and low-rank Hankel matrix reconstruction (HMR)
\cite{Cai2015Robust-Recovery} to estimate time delays or frequencies from the linear measurements.
For estimation from subsampled observations, we also consider algorithms such as atomic soft
thresholding (AST) for atomic norm minimization \cite{bhaskar2013atomic, tang2013compressed,
Chi2013Compressive-Rec, tang2013atomic} and enhanced matrix completion (EMC)
\cite{Chen2011Robust-Spectral}. All convex optimization methods (AST, EMC, HMR) are implemented
using the CVX package \cite{Grant2014CVX:-Matlab-Sof}, and the output time-domain signals are then
input to the root MUSIC algorithm for line spectral estimation. For time delay estimation, these
algorithms are modified according to (\ref{equation:modified}) to perform indirect time delay
estimation.  In the following experiments, the signal length is set for both applications to $N =
101$ due to the computational burden of the optimization-based methods. Table~\ref{table:algorithms}
illustrates the difference between the algorithms considered in these experiments.

\begin{table}[t]
\renewcommand{\arraystretch}{1.3}
\caption{\small \sl Comparison of algorithms for sparsity-based parameter estimation}
\label{table:algorithms}
\centering
\begin{tabular}{|m{2.2cm}|m{3cm}|m{2cm}|m{3cm}|m{2.5cm}|}
\hline
\bf Algorithm & \bf Estimation Problem & \bf Observation Model & \bf Performance Conditions & \bf Guarantee \\ 
\hline
AST {\footnotesize \cite{bhaskar2013atomic, tang2013compressed,Chi2013Compressive-Rec,
tang2013atomic}} & line spectrum estimation & subsampled signals & minimum separation, fast decay in
        correlation function & exact parameter estimation \\
\hline
BSP/BISP {\footnotesize \cite{fyhn2013compressive, duarte2013spectral, fyhn2013spectral}} &
arbitrary, with optional interpolation & arbitrary & maximum allowed coherence & exact signal recovery \\ 
\hline
EMC {\footnotesize \cite{Chen2011Robust-Spectral}} & line spectrum estimation & subsampled signals
& minimum separation & exact signal recovery\\
\hline
HMR {\footnotesize \cite{Cai2015Robust-Recovery}} & line spectrum estimation, NMR spectroscopy
estimation &
linear measurements & Gaussian measurement matrix &
exact signal recovery \\ 
\hline
AMP {\footnotesize \cite{Hamzehei2015Compressive-Par, Hamzehei2016Compressive-Dir}} & arbitrary with statistical estimation from noisy observations  & linear measurements & Gaussian measurement matrix, sufficient denoiser performance & recovery prediction via state evolution \\
\hline
CSP/CISP & arbitrary, with optional interpolation & arbitrary & minimum separation, fast decay in
correlation function & approximated parameter estimation \\
\hline
\end{tabular}
\end{table}

\begin{figure}[t]
\centering
\includegraphics[width = 0.4\textwidth]{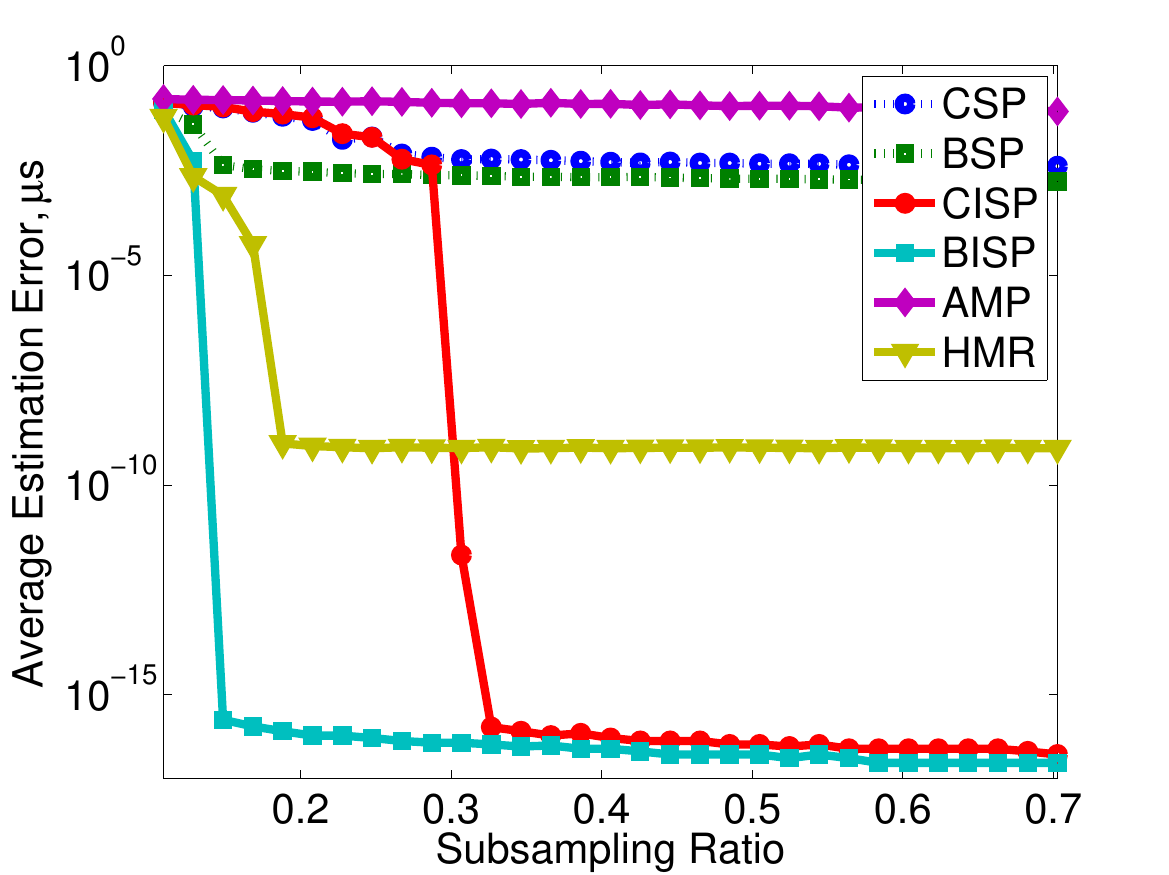}
\includegraphics[width = 0.4\textwidth]{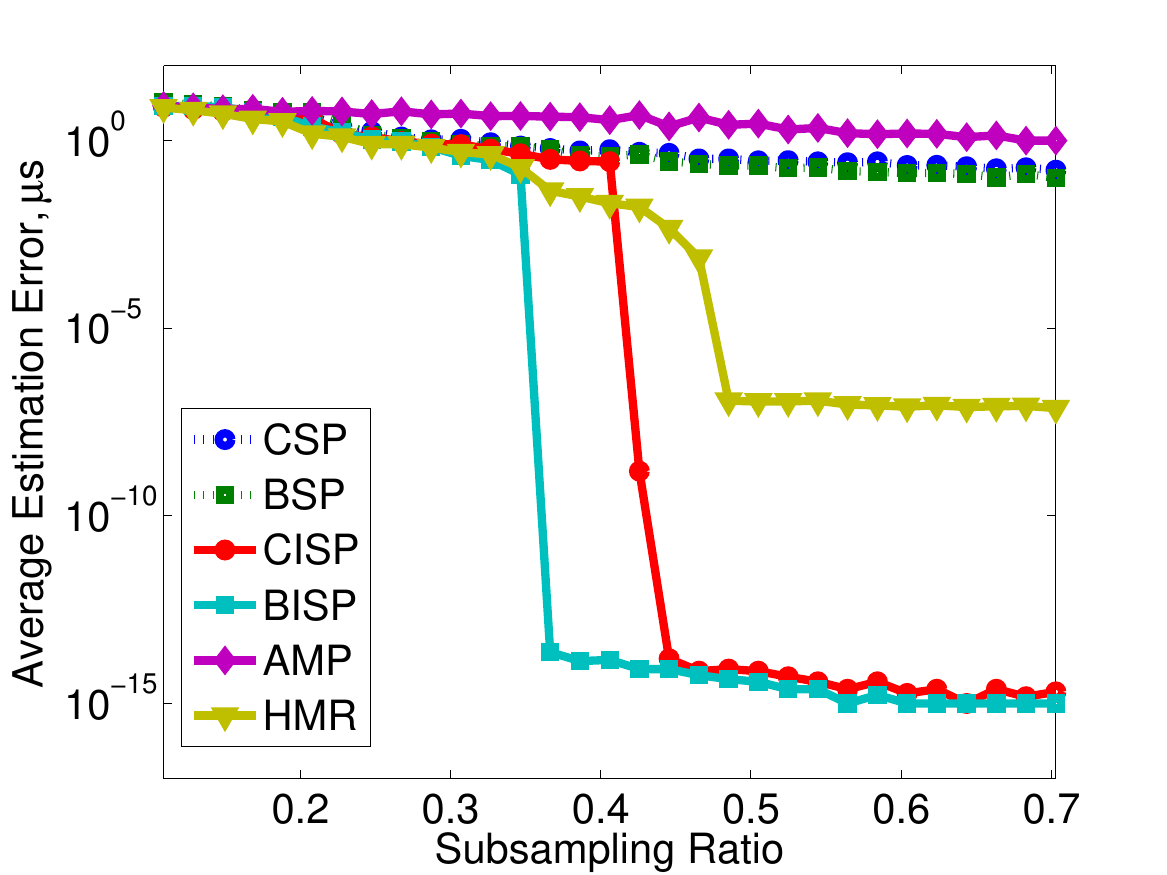}
\caption{\small \sl Performance for the time delay estimation problem, as measured by the average
parameter error, when (left) linear measurements and (right) subsampled signals are observed.}
\label{figure:TDE1}
\vspace{-5mm}
\end{figure}

{\bf Time Delay Estimation:} Our first experiment tests the estimation performance of the algorithms
listed in this section on $1000$ independent randomly-generated time delay estimation problems. The
chirp signal has frequency sweep $f_a = 4~\text{MHz}$ and sampling frequency $f_s = 10~\text{MHz}$.
Thus, the duration of the whole chirp signal is $N T_s = 10.1~\mu\text{s}$. The $K=4$ unknown time
delays are generated randomly such that the minimum separation $\zeta = 0.05~\mu\text{s}$. The
parametric dictionary used for CSP, CISP, BSP and BISP contains the signal observations for the
sampling set with sampling step $\Delta = 0.01~\mu\text{s}$. The threshold value for CSP and CISP is
set to $t = 0$ (i.e., no thresholding takes place). The maximum allowed coherence for BSP and BISP
is chosen as $\mu = 0.01$, which was found to give the best estimation performance over a grid
search.  Figure~\ref{figure:TDE1} shows the average estimation error as a function of the
compression rate $\kappa = M/N$ when noiseless linear measurements and subsampled signals are
observed. The linear measurements are obtained via a measurement matrix with entries drawn i.i.d.\
from the standard Gaussian distribution. The results indicate that CSP can estimate the time delays
with error below the sampling step $\delta$ when $\kappa \ge 0.3$ for both types of observations,
which matches the performance of its band-exclusion counterparts. However, while BSP requires
setting the maximum allowed coherence $\mu$ carefully via grid search, CSP can achieve similar
performance without relying on thresholding. Due to the limitation of discretization, both CSP and
BSP perform worse than HMR, which uses convex optimization to estimate off-grid time delays
precisely. Armed with polar interpolation, both BISP and CISP can easily reach accurate estimation.
We believe that the performance of HMR can increase with the use of more precise convex optimization
solvers; however, this would also increase the estimation complexity. The poor performance of AMP is
likely due to the fact that the measurement matrix derived from (\ref{equation:modified}) does not
allow for the signal estimate to be modeled statistically as the true signal with additive white
Gaussian noise \cite{Hamzehei2015Compressive-Par}. Additionally, AST and EMC cannot estimate the
time delay from subsampled signals since the process of changing the time delay estimation to
frequency estimation is possible only for linear measurements.

\begin{figure}[t]
    \centering
\includegraphics[width = 0.4\textwidth]{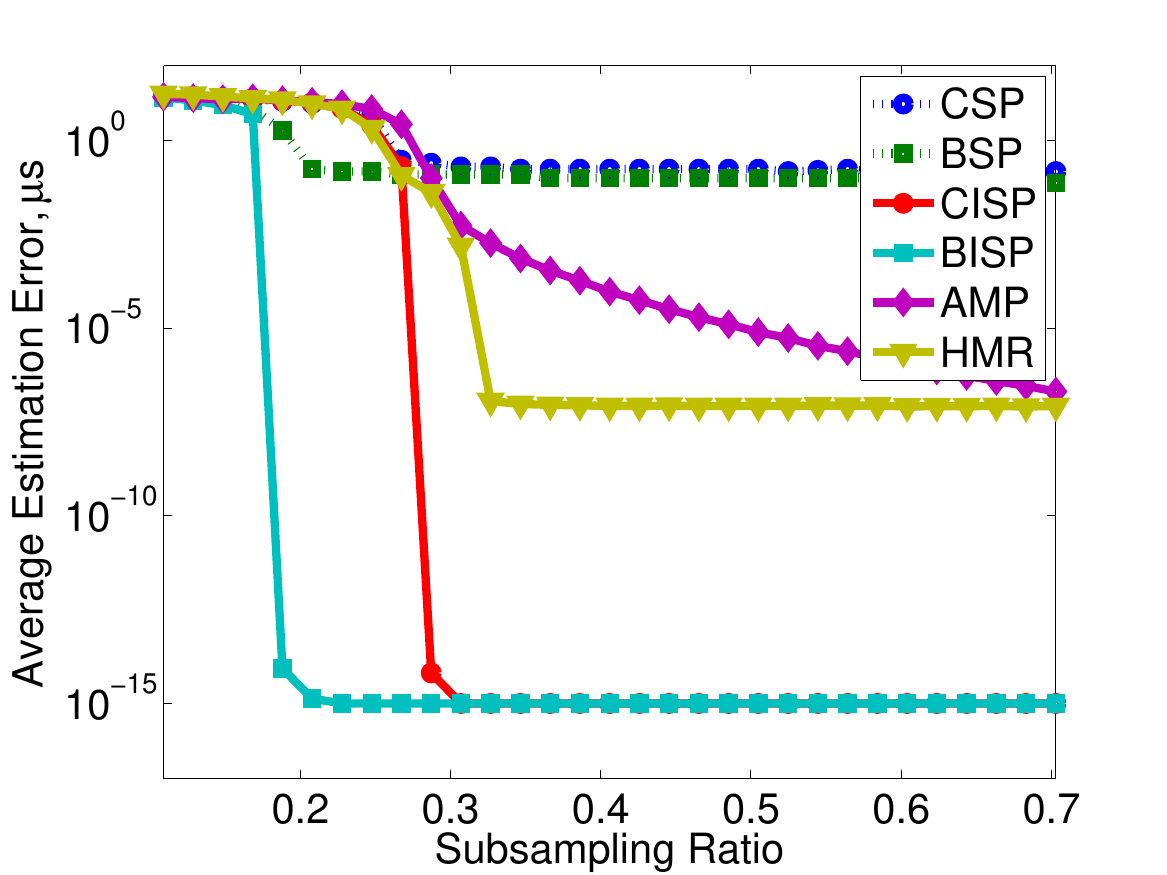}
\includegraphics[width = 0.4\textwidth]{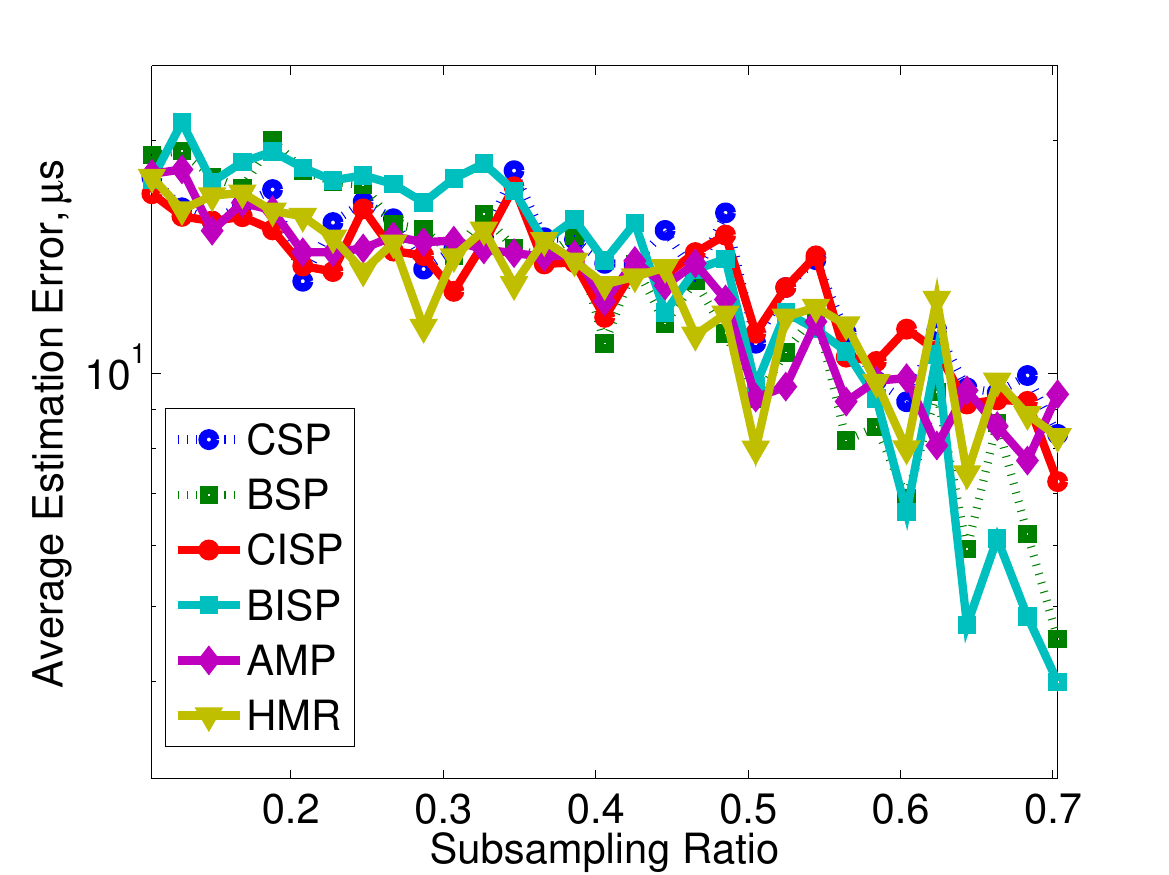}
\caption{\small \sl Performance for the impulse delay estimation problem, as measured by the average
parameter error, when (left) linear measurements and (right) subsampled signals are observed.}
\label{figure:TDE2}
\end{figure}

When the duration of the chirp signal is decreased, the Gram matrix $G$ in (\ref{equation:modified}) converges towards a scaled identity
matrix. Thus, the measurement matrix for the resulting frequency estimation problem more closely 
matches that of the original delay estimation problem. Figure \ref{figure:TDE2} presents the 
average estimation error for an impulse delay estimation as the function of the compression rate 
$\kappa$. The good performance of AMP in estimation from linear measurements demonstrates 
that the necessary properties of measurement matrix remain. It is easy to see that when the
observations correspond to a subsampled signal, it will be very difficult to estimate the impulse time
delays.

{\bf Frequency Estimation:} In our next experiment, we repeat the first experiment on frequency estimation with $1000$
independent randomly drawn signals. The $K=4$ unknown frequencies have minimum separation $\zeta = 0.1 \text{Hz}$. The parametric dictionary used
for CSP, CISP, BSP and BISP contains the signal observations for the sampling set with sampling 
step $\Delta = 0.1 \text{Hz}$. The threshold value for CSP and CISP and the maximum
allowed coherence for BSP and BISP are required to be $t = 0.4$ and $\mu = 0.1$ to have good
performance, according to Theorem~\ref{theorem:performance} and a grid search, respectively.

\begin{figure}[t]
\centering
\includegraphics[width = 0.4\textwidth]{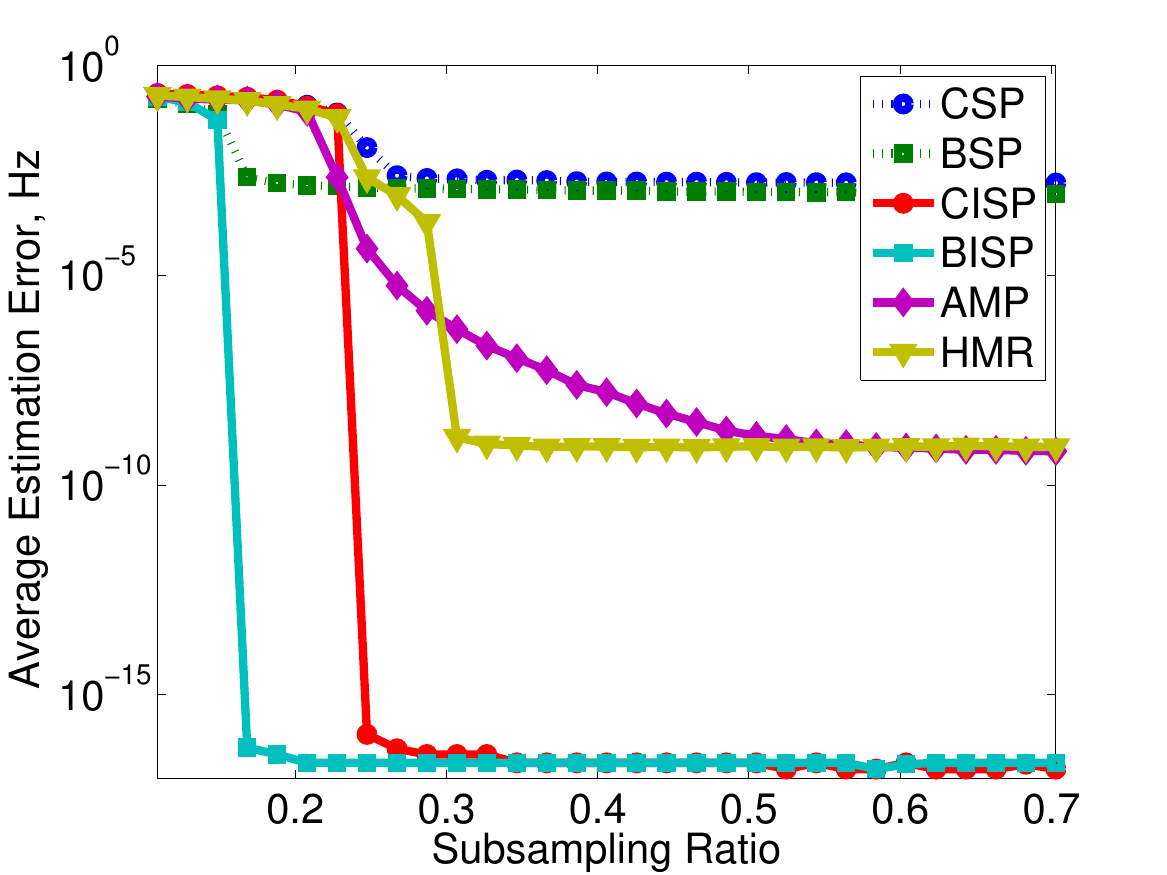}
\includegraphics[width = 0.4\textwidth]{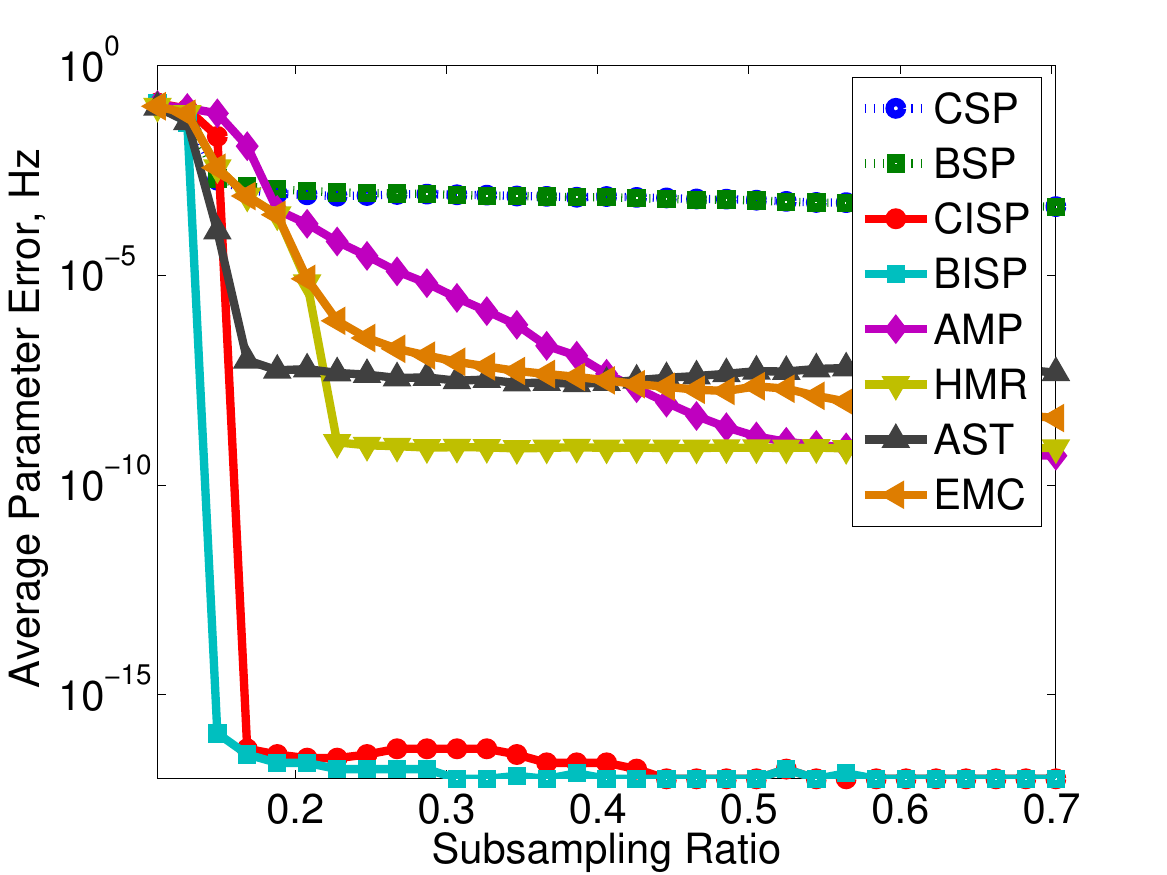}
\includegraphics[width = 0.4\textwidth]{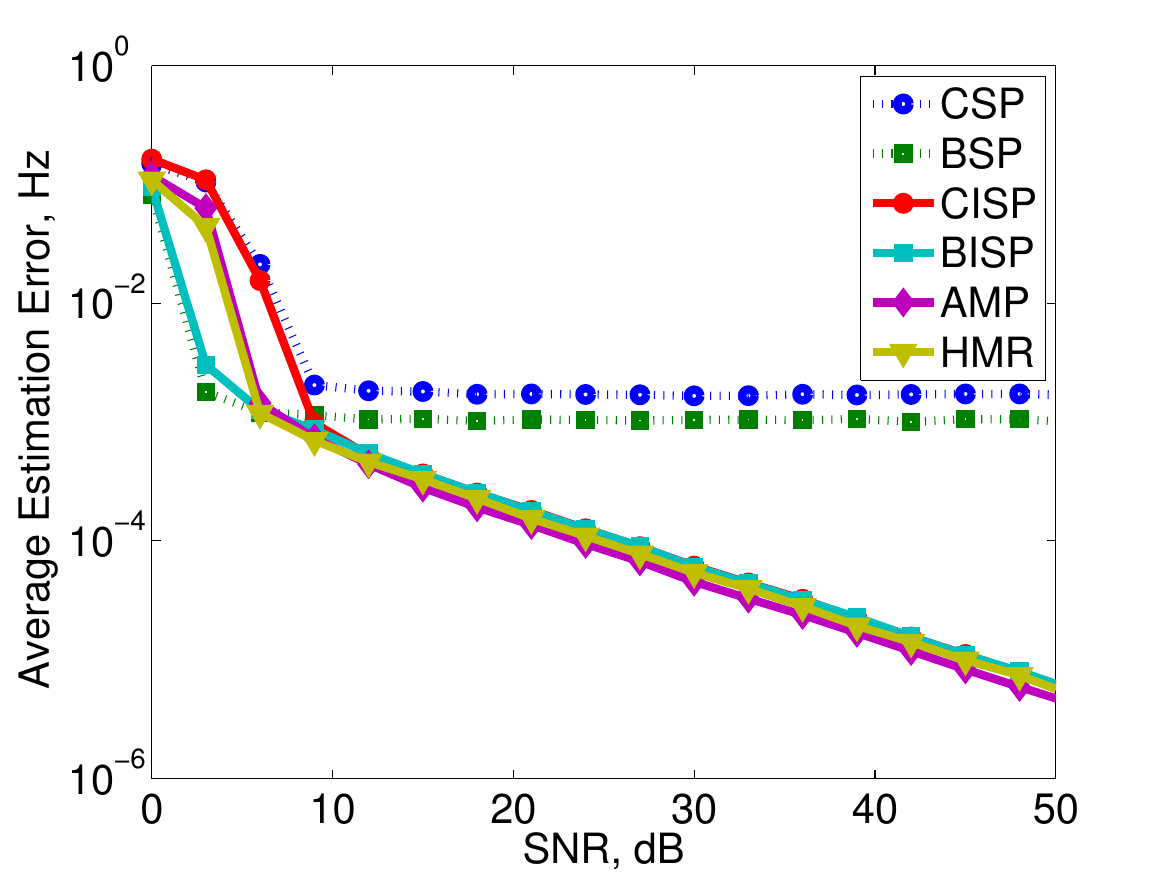}
\includegraphics[width = 0.4\textwidth]{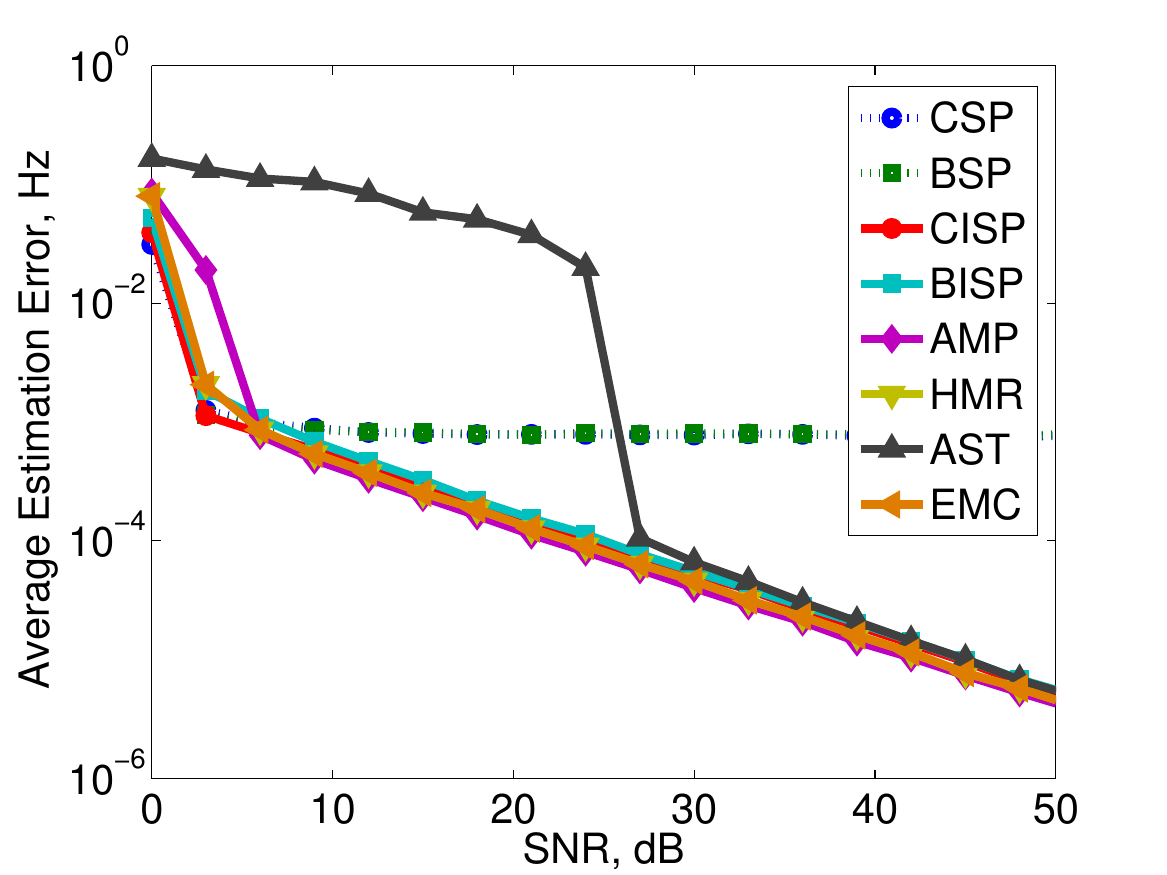}
\caption{\small \sl Performance of \CPE~for the frequency estimation, as measured by the average parameter
error, as a function of (top) the \CS~compression rate $\kappa = M/N$ with noiseless measurements
and (bottom) the measurement SNR with $\kappa = 0.4$, under (left) linear measurements and (right)
subsampled observations.}
\label{figure:FE}
\vspace{-5mm}
\end{figure}

Similarly as in the first experiment, Figure \ref{figure:FE} shows the average estimation error as a
function of the compression rate and the SNR. The conclusion is similar: armed with polar
interpolation, CISP achieves accurate estimation when compression is high enough and CISP shows
similar noise robustness as other algorithms; we found that the trends shown in the figure extend to SNRs of at least 90~dB. Though CSP needs the thresholding value to estimate
the frequencies correctly, we can always choose a proper value according to Theorem
\ref{theorem:performance}.

\begin{figure}[t]
\centering
\includegraphics[width = 0.45\textwidth]{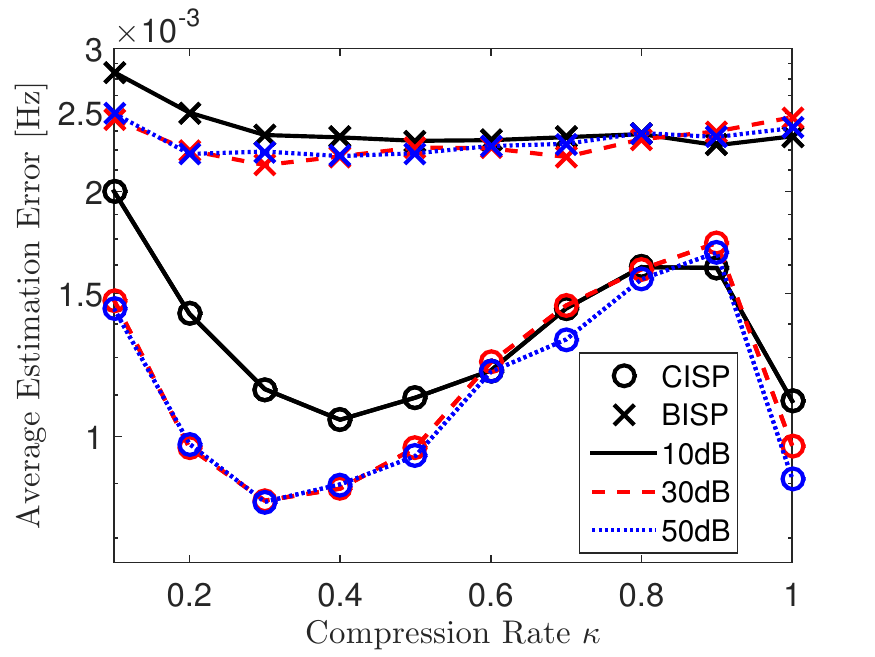}
\caption{\small \sl Performance of \CPE~ algorithm for the real \FE~problem. Dash-dot, dashed,
dotted lines represent the average relative estimation errors when SNR = 10, 30, 50 dB, and lines
with circle and cross are errors of CISP and BISP.}
\label{figure_real}
\vspace{-5mm}
\end{figure}

In the last experiment, we apply our proposed clustering-based algorithm for \CPE~with a real-world
signal. The lynx signal has $N=114$ samples and is used to test the performance of line spectral
estimation algorithms in \cite{stoica2005spectral}. It is well approximated by a sum of complex
sinusoids with small minimum separation distance and large dynamic range among the component
magnitudes. We increase the size of the signal by a factor of 10 using interpolation and obtain CS
measurements of the resulting signal with a random matrix for various compression rates under
several levels of measurement noise. The maximum allowed coherence in BISP is set to $\nu = 0.2$
after a grid search to optimize the algorithm's performance, while the threshold level in CISP is
set to $t = 0$. Figure~\ref{figure_real} shows the average relative estimation error between the
estimates from CISP and BISP and those obtained from root MUSIC (a line spectral estimation
algorithm with high accuracy \cite{stoica2005spectral}) when applied to the full length signal. The
results show that CISP without thresholding has closer performance to root MUSIC than the best
configuration of BISP.

\section{Discussion and Conclusions}
\label{section:conclusion}

In this paper, we have introduced and analyzed the relationship between the \EMD~applied to
\PD~coefficient vectors and the \PEE~obtained from sparse approximation methods applied to the
\PD~representation. We also leveraged the relationship between \EMD-based sparse approximation and
$K$-median clustering algorithms in the design of new \CPE~algorithms. Based on the relationship
between the \EMD~and the \PEE, we have analytically shown that the \EMD~between \PD~coefficient
vectors provides an upper bound of the \PEE~obtained from methods that use \PDs~and \EMD.
Furthermore, we leveraged the connection between \EMD-sparse approximation and $K$-median
clustering, to formulate new algorithms that employ sparse approximation in terms of the \EMD; we
then derived three theoretical results that provide performance guarantees for \EMD-based parameter
estimation under certain requirements for the signals observed, in contrast to existing work that
does not provide such guarantees. Our experimental results show the validation of our analysis in
several practical settings, and provides methods to control the effect of coherence, compression,
and noise in the performance of \CPE.

In our new \CPE~algorithms, we use $K$-median clustering rather than the more predominant $K$-means
clustering to obtain the sparse approximation or the local maxima of the proxy. The main difference
between $K$-median clustering and $K$-means clustering is that their criteria in
(\ref{equation:generalobject}) are different: the former uses the Manhattan distance and the latter
uses the squared Euclidean distance. This difference makes $K$-median clustering more robust to
noise and outliers and prevents $K$-means clustering from being able to return the \EMD-optimal
sparse approximation in general.

Our experiments have shown that our clustering-based algorithms for \CPE~can achieve the same
performance as those based on band exclusion. Though both methods use additional parameters to
improve the performance, the clustering method is preferable as it does not need to rely on
parameter tuning under a noiseless setting and for simpler problems, while the band exclusion is
highly dependent on the allowed coherence in all cases. The threshold level is needed only in cases
where the estimation problem is particularly ill-posed, \CS~has been heavily used, measurement noise
is present, or in other cases where the correlation function decays slowly. As shown in Figure
\ref{figure:TDE1}, the clustering method without thresholding has the same performance as the
band-exclusion method with optimally tuned maximum coherence. Interested readers can refer to
\cite{mo2013compressive}, where we have further studied the different sensitivities of these two
methods on the additional parameters.

The results we obtained show the very different performance level of convex optimization-based 
methods on time delay estimation and frequency estimation, which is due to the fact that all convex 
methods can not be applied on time delay estimation directly. In contrast, the advantage of 
\PD-based parameter estimation is that it can be utilized on almost all parameter estimation 
problems, as they only require that a \PD~can be obtained to yield sparse representation for the 
relevant signals. Additionally, the convex methods depend on convex optimization solver to
recover signals, which have high computation cost. In the second part of numerical experiments, 
we have to set the length of the signal to be a small value otherwise the CVX package is not able to
solve the proposed optimization problem. Instead, the essential part to the \PD-based parameter
estimation is the sparse approximation operator used in the algorithm, which usually can be much
simpler. Furthermore, as shown in Table~\ref{table:algorithms}, the observation model of \PD-based
parameter estimation can be arbitrary: the observations can be the linear measured or
subsampled signals, while existing approaches focus on a single observation model.

\section*{Acknowledgements}

We thank Armin Eftekhari and Michael Wakin for helpful comments and for pointing us to
\cite{eftekhari2014new-analysis}.

\appendix

\section{Proof of Theorem \ref{theorem:isometric}}
\label{chapter:isometryproof}

Assume that the vectors $c$ and $\hc$ have the same $\ell_1$ norm, so that the standard definition
of the EMD applies. Let $f^* = \left[ f_{1, 1}^*, f_{1, 2}^*, \dots, f_{1, K}^*, f_{2, 1}^*, \dots,
    f_{K, K}^*
\right]^T$ be the vector that solves the optimization problem (\ref{equation:emd}), $g^*$ be the
similarly-defined binary vector that solves the optimization problem (\ref{equation:pee}), and $z =
f^*-c_{\min}g^*$ is a flow residual. Similarly, let $d$ and $t$ be the similarly-defined vectors
collecting all ground distances $d_{i, j}$ and $t_{i, j}$. Then from (\ref{equation:pee}) and
(\ref{equation:emd}), we have
\begin{align}
\operatorname{EMD} \left( c,\widehat{c} \right) = d^Tf^{*}  = d^T \left( c_{\min}g^{*} + z \right) =
\frac{c_{\min}}{\Delta} t^Tg^{*} + d^T z = \frac{c_{\min}}{\Delta} \operatorname{PEE} \left(
\theta,\widehat{\theta} \right) + d^T z.
\label{equation:errorrelationship}
\end{align}
Note that the first term in (\ref{equation:errorrelationship}) is the value of the objective
function in the optimization problem (\ref{equation:emd}) when all entries of both $c$ and $\hc$
have magnitude $c_{\min}$, i.e., when $z = 0$. The second term $d^Tz$ corresponds to the
contribution to the objective function due to magnitudes that are larger than $c_{\min}$. We show
now that this second term is nonnegative.

When the magnitude of the entry of $c$ corresponds to $\hp _i$ increases from its baseline value of $c_{\min}$
to $c_i$, at least one of the outgoing flows $f_{i, j}$ have to increase. This implies that the
corresponding flow $f^*_{i, j} \ge c_{\min} g^{*}_{i, j}$, i.e., the residual $z_{i, j} =
f^*_{i, j}-c_{\min}g^*_{i, j} \ge 0$.  Thus, the entries $r$ are all nonnegative, and $d^Tr \geq 0$.
Then one can rewrite (\ref{equation:errorrelationship}) as $\text{EMD}(c,\widehat{c}) \geq
\frac{c_{\min}}{\Delta} \text{PEE}\left( \theta,\widehat{\theta} \right)$.

To end, we consider the case when $\| c \| _1 \ne \left \| \hc \right \| _1$, where we have to add additional entries to the vectors: $c _{N+1} = c_{\min} + \max \left\{ 0, \left\| \hc \right\| _1 - \left\| c
\right\| _1 \right\}$ and $\hc _{N+1} = c_{\min} + \max \left\{ 0, \left\| c \right\| _1 - \left\| \hc
\right\| _1 \right\}$. The additional estimated parameters due to those additional entries are both
$\theta_m$ since the indices of the additional entries are the same. 
The resulting $K+1$-sparse vectors with additional entries are $c^* = \left[ c_1, c_2, \dots, c_N, c_{N+1} \right]$ 
and $\hc^* = \left[ \hc_1, \hc _2, \dots, \hc_N, \hc _{N+1} \right]$; these two new vectors have matching $\ell_1$ norms, $\|c^*\|_1 = \|\hc^*\|_1$, and the resulting estimated 
parameters are $\theta ^* = \left[ \theta_1, \theta _2, \dots, \theta_K, \theta_m \right]$ and 
$\hp ^* = \left[\hp_1, \hp _2, \dots, \hp_ K, \theta_m \right]$, where $\theta_m$ is an (arbitrary) parameter value assigned to the newly added $N+1$-index entry. Therefore, we get 
$\text{PEE} \left( \theta, \hp \right) = \text{PEE}
\left( \theta ^*, \hp ^* \right)$, due to a match between the added parameter values $\theta _m$ in 
$\theta ^*$ and in $\hp ^*$. 

Next, we consider the effect of the mismatch of the $\ell_1$ norms of the vectors $c$ and $\hc$ in the computation of $\text{EMD} \left(c, \hc \right)$ and $\text{EMD} \left( c ^*, \hc ^* \right)$.
We can see that $\text{EMD} \left(c, \hc \right)  \ge \text{EMD} \left( c ^*, \hc ^* \right)$: when we compute $\text{EMD} \left(c,\hc \right)$, the $\ell_1$ norm mismatch penalty in the objective function of (\ref{equation:emd}) -- as defined in Footnote 1 and involving the flows $f_{i,N+1}$ and $f_{N+1,j}$ from/to the entries $c _{N + 1}, \hc_{N+1}$ to/from any $c_j,\hc _j$ -- is no less than 
the actual contribution of the flows from/to the new entries $c^*_{N+1},\hc^*_{N+1}$ to/from the
existing entries $c^*_j,\hc^* _j$ in computing $\text{EMD} \left( c ^*, \hc ^* \right)$, as those
distances $d_{i,N+1}$ and $d_{N+1,j}$ are all less than or equal to $N$. Putting this together with
the equivalence in PEE and Theorem~\ref{theorem:isometric} when applied to vectors with matching
$\ell_1$ norms,we obtain that 
$\frac{c _{\min}}{\Delta} \text{PEE} \left( \theta, \hp \right) = \frac{c_{\min}}{\Delta} \text{PEE} \left( \theta ^*, \hp ^* \right) \le \text{EMD} \left( c^*, \hc ^*\right) \le \text{EMD} \left( c, \hc \right)$.

\section{Proof of Theorem \ref{theorem:generalclustering}}
\label{chapter:clusteringproof}

As the sampling step of the parameter $\Delta \to 0$, the proxy defined as
(\ref{equation_proxy}) becomes a continuous function such that
\begin{equation}
v(\omega) = \sum_{i=1}^K c_i \lambda(\omega - \theta_i),
\label{equation:continuousproxy}
\end{equation}
for all $\omega \in \Thetabar$. Additionally, the balanced weight properties around the cluster
centroid $\widehat{\theta}_i$, as defined in (\ref{equation:balanceweight}), reduces in the asymptotic case to
\begin{equation}
\int_{p \in C_j, p \le \widehat{\theta}_i} w(p) \der p = \int_{p \in C_j, p \ge \widehat{\theta}_j} w(p) \der p,
\label{equation:continuousbalance}
\end{equation}
where $p$ is the position function and $w$ is the weight function. Furthermore, the cumulative correlation function converges to
\begin{equation}
\Lambda(\theta) = \int_{-\infty}^{\theta} \lambda(\omega) \der\omega.
\end{equation}

Without loss of generality, if $\theta_{\min} = \min \left( \Thetabar \right)$ and $\theta_{\max} =
\max \left( \Thetabar \right)$, assume that parameter values are sorted so that
\begin{equation}
\theta_{\text{min}} +\epsilon \leq \theta_1<\theta_2<\dots < \theta_K \leq \theta_{\text{max}} -\epsilon.
\end{equation}

When the entries of the proxy $v$ are clustered into $K$ groups according to the centroids
$\hp_1, \hp_2, \dots, \hp_K$, as shown in Algorithm
\ref{algorithm:clustering}, the point $\left( \hp_i + \hp_{i+1} \right)/2$ is the upper
bound for $i ^{\text{th}}$ cluster and the lower bound for $(i+1) ^ {\text{th}}$ cluster, since that 
value/location has the same distance to the two centroids around it. We will show how large the 
minimum separation $\zeta$ and minimum off-bound distance $\epsilon$ need to be such that 
the maximum estimation error is $\sigma$, i.e. $\max_k \left| \hp_k - \theta_k \right| \le \sigma$.

We first consider the $k ^{\text{th}}$ cluster with centroid $\hp_k$, where $k = 2, 3, \dots, K-1$.
The $k ^{ \text{ th } }$ cluster includes the parameter range $\left[ \left( \hp _{k-1} + \hp _ k \right) / 2, \left( \hp _k + \hp
_{k+1} \right) / 2 \right]$. According to the weight balance property (\ref{equation:continuousbalance}),
we need for the proxy function in (\ref{equation:continuousproxy}) to have the same sum over the range
$\left[ \left( \hp_{k-1} + \hp_k \right) / 2 , \hp_k \right]$ and $\left[ \hp_k, \left( \hp_k +
\hp_{k+1} \right) / 2 \right]$, i.e.,
\begin{align}
\int_{\frac{\hp_{k-1}+\hp_k}{2}}^{\hp_k} v(\omega) \dw &=
\int_{\hp_k}^{\frac{\hp_k+\hp_{k+1}}{2}} v(\omega) \dw, \nonumber \\
2 \int_{-\infty}^{\hp_k} v(\omega) \dw &=
\int_{-\infty}^{\frac{\hp_{k-1}+\hp_k}{2}} v(\omega) \dw+ 
\int_{-\infty}^{\frac{\hp_k+\hp_{k+1}}{2}} v(\omega) \dw, \nonumber \\
2\sum_{i =1}^{K} c_i\Lambda \left(\hp_k-\theta_i \right) &=
\sum_{i = 1}^K c_i \Lambda\left(\frac{\hp_{k-1}+\hp_k}{2}-\theta_i\right)+
\sum_{i = 1}^K c_i \Lambda\left(\frac{\hp_k+\hp_{k+1}}{2}-\theta_i\right), \nonumber \\
2 c_k \Lambda \left( \hp_k - \theta_k \right) &= 
\sum_{i = 1}^K c_i \Lambda\left(\frac{\hp_{k-1}+\hp_k}{2}-\theta_i\right)+
\sum_{i = 1}^K c_i \Lambda\left(\frac{\hp_k+\hp_{k+1}}{2}-\theta_i\right)-
2\sum_{i \ne k} c_i\Lambda \left(\hp_k-\theta_i \right).
\label{equation_middlecondition}
\end{align}

Since $\hp_k - \theta_k \in \{ - \sigma , \sigma \}$ and $\theta_{k+1} - \theta_k \ge \zeta$, for $k =
2, 3, \cdots, K-1$, we obtain a lower bound of the term on the left hand side of
(\ref{equation_middlecondition}) by repeatedly using the fact that $\Lambda(\omega)$ is
nondecreasing:
\begin{align}
&\sum_{i = 1}^K c_i \Lambda\left(\frac{\hp_{k-1}+\hp_k}{2}-\theta_i\right)+
\sum_{i = 1}^K c_i \Lambda\left(\frac{\hp_k+\hp_{k+1}}{2}-\theta_i\right)-
2\sum_{i \ne k} c_i\Lambda \left(\hp_k-\theta_i \right) \notag \\
= &\sum_{i = 1} ^{k-1} c_i \Lambda\left(\frac{\hp_{k-1}+\hp_k}{2}-\theta_{k-1}\right) +
\sum_{i = k} ^{K} c_i \Lambda\left(\frac{\hp_{k-1}+\hp_k}{2}-\theta_K\right) + \sum_{i = 1} ^{k} c_i
\Lambda\left(\frac{\hp_{k}+\hp_{k+1}}{2}-\theta_{k}\right) \notag \\
& \qquad + \sum_{i = k+1} ^{K} c_i \Lambda\left(\frac{\hp_{k}+\hp_{k+1}}{2}-\theta_K\right) - 2 \sum_{i =
1}^{k-1} c_i\Lambda \left(\hp_k-\theta_1 \right) - 2\sum_{i = k+1}^K c_i\Lambda
\left(\hp_k-\theta_{k+1} \right) \notag \\
\ge& \sum_{i = 1} ^{k-1} c_i \Lambda\left(\frac{\hp_{k-1} - \theta_{k-1}+\hp_k - \theta_k + \theta_k
- \theta_{k-1}}{2}\right) + \sum_{i = k} ^{K} c_i \Lambda\left(-\infty\right) \notag \\
& \qquad +\sum_{i = 1} ^{k} c_i \Lambda\left(\frac{\hp_{k} - \theta_{k}+\hp_{k+1} - \theta_{k+1} +
\theta_{k+1} - \theta_{k}}{2}\right) + \sum_{i = k+1} ^{K} c_i \Lambda\left(-\infty\right) - 2
\sum_{i = 1}^{k-1} c_i\Lambda \left(\infty \right) \notag \\
& \qquad - 2\sum_{i = k+1}^K c_i\Lambda \left(\hp_k-\theta_{k} + \theta_{k} - \theta_{k+1} \right)
\notag \\
\ge& \sum_{i = 1}^{k-1} c_i \Lambda \left( \frac{\zeta}{2} - \sigma \right) + \sum_{i = 1}^{k} c_i
\Lambda \left( \frac{\zeta}{2} - \sigma \right) - 2 \sum_{i = 1}^{k-1} c_i\Lambda \left(\infty \right) -
2\sum_{i = k+1}^K c_i\Lambda \left( -\frac{\zeta}{2} + \sigma \right) \notag \\
\ge& \sum_{i = 1}^{k-1} c_i \Lambda \left( \frac{\zeta}{2} - \sigma \right) + \sum_{i = 1}^{k} c_i
\Lambda \left( \frac{\zeta}{2} - \sigma \right) - 2 \sum_{i = 1}^{k-1} c_i\Lambda \left(\infty
\right) - 2\sum_{i = k+1}^K c_i \left( \Lambda (\infty) - \Lambda \left( \frac{\zeta}{2} - \sigma
\right) \right) \notag \\
\ge& - 2 \sum_{i \ne k }  \frac{c_i}{c_k}  c_k \left( \Lambda \left( \infty \right) - \Lambda \left(
\frac{\zeta}{2} - \sigma \right) \right) + c_k \Lambda \left( \frac{\zeta}{2} - \sigma \right) \nonumber\\
\ge& -
2(K-1)r c_k \left( \Lambda \left( \infty \right) - \Lambda \left( \frac{\zeta}{2} - \sigma \right)
\right) + c_k \Lambda \left( \frac{\zeta}{2} - \sigma \right) \notag \\
\ge & \left( 2(K-1) r + 1 \right) c_k \Lambda \left( \frac{\zeta}{2} - \sigma \right) - 2(K-1)r c_k
\Lambda(\infty).
\end{align}
So the lower bound of the term on the left hand side of (\ref{equation_middlecondition}) is
\begin{align}
2 \Lambda \left( \hp_k - \theta_k \right) \ge \left( 2(K-1) r + 1 \right) \Lambda \left(
\frac{\zeta}{2} - \sigma \right) - 2(K-1)r \Lambda(\infty).
\end{align}
Similarly, one can obtain the following upper bound of term on the left hand side of (\ref{equation_middlecondition}):
\begin{align}
2 \Lambda \left( \hp_k - \theta_k \right) \le \left(2(K-1)r+2\right) \Lambda(\infty) - \left( 2(K-1)
r + 1 \right) \Lambda \left( \frac{\zeta}{2} - \sigma \right).
\end{align}
Thus, if $\zeta$ satisfies
\begin{align}
\Lambda \left( \frac{\zeta}{2}- \sigma \right) \ge \Lambda(\infty) \left(1 - \frac{\Lambda(\sigma) /
\Lambda(0) -1 }{2(K-1)r + 1} \right),
\label{equation:minsep}
\end{align}
then the estimation error satisfies
\begin{align}
2 \Lambda \left( \hp_k - \theta_k \right) &\ge \left( 2(K-1) r + 1 \right) \Lambda \left(
\frac{\zeta}{2} - \sigma \right) - 2(K-1)r \Lambda(\infty) \notag \\
&\ge \left( 2(K-1) r + 1 \right) \Lambda(\infty) - 2\Lambda(\sigma) + \Lambda(\infty) - 2(K-1)r
\Lambda(\infty) \notag \\
&\ge 2\Lambda(\infty) - 2\Lambda(\sigma)\ge 2\Lambda(-\sigma),
\end{align}
and
\begin{align}
2 \Lambda \left( \hp_k - \theta_k \right) &\le \left(2(K-1)r+2\right) \Lambda(\infty) - \left(
2(K-1) r + 1 \right) \Lambda \left( \frac{\zeta}{2} - \sigma \right) \notag \\
&\le \left(2(K-1)r+2\right) \Lambda(\infty) - \left( 2(K-1) r + 1 \right) \Lambda (\infty) +
2\Lambda(\sigma) - \Lambda(\infty) \notag \\
&\ge 2\Lambda(\sigma),
\end{align}
which implies $-\sigma \le \hp_k - \theta_k \le \sigma$ for $k = 2, 3, \cdots, K-1$.\par

Next, we consider the first cluster with centroid $\hp_1$, which includes the parameter range
$\left[ \theta_{\min} , \left( \hp _1+\hp _2 \right) / 2 \right]$. From the weight balance property,
we have
\begin{align} 
    \int _{ \theta_{\min} } ^{ \hp _1 } v ( \omega ) \dw & = \int _{ \hp _1 } ^{ \frac{ \hp _1 + \hp
    _2}{2} } v ( \omega ) \dw, \nonumber \\
    2 \int _{ -\infty } ^{ \hp_1 } v( \omega ) \dw &= \int _{ -\infty } ^{ \theta _{ \min } } v(
    \omega ) \dw + \int _{ -\infty } ^{ \frac{ \hp_1 + \hp_2 }{2} } v( \omega ) \dw, \nonumber \\
    2\sum_{i = 1}^K c_i \Lambda \left( \hp_1-\theta_i \right) &= \sum_{i = 1}^K c_i
    \Lambda \left( \theta_{\min}-\theta_i \right)+ \sum_{i = 1}^K c_i
    \Lambda \left( \frac{\hp_1+\hp_2}{2} - \theta_i \right), \nonumber \\
    2c_1 \Lambda \left( \hp_1-\theta_1 \right) &= \sum_{i = 1}^K c_i \Lambda \left( \theta_{\min}-\theta_i
    \right) + \sum_{i = 1}^K c_i \Lambda \left( \frac{\hp_1+\hp_2}{2} - \theta_i \right) - 2\sum_{i
    = 2}^K c_i \Lambda \left( \hp_1 - \theta_i \right).
    \label{equation:firstbalance}
\end{align}
If $\epsilon$ satisfies 
\begin{align}
\Lambda(\epsilon) \ge \Lambda(\infty) \left(1 - \frac{\Lambda(\sigma) / \Lambda(0) - 1}{2Kr} \right),
\label{equation:minbound}
\end{align}
then we have the following result from (\ref{equation:firstbalance}) by using the relationship
(\ref{equation:minsep}):
\begin{align}
2 \Lambda \left( \hp_1-\theta_1 \right) & = \sum_{i = 1}^K \frac{c_i}{c_1} \Lambda \left(
    \theta_{\min}-\theta_i \right) +
    \sum_{i = 1}^K \frac{c_i}{c_1} \Lambda \left( \frac{\hp_1+\hp_2}{2} - \theta_i \right) - 2
    \sum_{i = 2}^K \frac{c_i}{c_1} \Lambda \left( \widehat{\theta}_1-\theta_i \right) \notag \\
& \le \sum_{i = 1}^K \frac{c_i}{c_1} \Lambda \left( \theta_{\min} - \theta_1 \right) + \Lambda \left(
\frac{\hp_1+\hp_2}{2} - \theta_1 \right) + \sum_{i = 2}^K \frac{c_i}{c_1} \Lambda \left(
\frac{\hp_1+\hp_2}{2}-\theta_2 \right) - 2 \sum_{i = 2}^K \frac{c_i}{c_1} \Lambda \left( \hp_1 -
\theta_K \right) \notag \\
& \le Kr \Lambda ( -\epsilon ) + \Lambda ( \infty ) + (K-1)r \Lambda \left( -\frac{\zeta}{2} +
\sigma \right) \notag \\
& \le Kr \left( \Lambda(\infty) - \Lambda ( \epsilon ) \right) + \Lambda ( \infty ) + (K-1)r \left(
\Lambda ( \infty ) - \Lambda \left( \frac{\zeta}{2} - \sigma \right) \right) \notag \\
& \le Kr \frac{ 2 \Lambda (\sigma) - \Lambda(\infty)}{ 2Kr} + \Lambda(\infty) + (K-1) r \frac{2
\Lambda (\sigma) - \Lambda(\infty) }{ 2 (K - 1) r + 1} \notag \\
& \le 2\Lambda(\sigma)
\end{align}
It can be similarly shown for the last cluster centroid that $2\Lambda \left( \hp_K - \theta_K
    \right) \ge 2 \Lambda(-\sigma)$.

In summary, when the minimum separation $\zeta$ and off-bound distance $\epsilon$ satisfy equation
(\ref{equation:minsep}) and (\ref{equation:minbound}), respectively, all estimates satisfy $\left| \hp _k
- \theta _k \right| \le \sigma, k = 1, 2, \dots, K$, and hence $\text{PEE} \left( \hp, \theta
\right) \le K \sigma$

\section{Proof of Theorem \ref{theorem:performance}}
\label{chapter:performanceproof}

When the redefined correlation function is $\lambda_{\Phi}(\omega) = \exp\left(-a |\omega| \right)$, the proxy function given in (\ref{equation_induced_proxy}) is
\begin{equation}
v (\omega) = \sum_{i=1}^K c_i \exp\left( - a \left|\omega - \theta_i \right| \right).
\end{equation}
Without loss of generality, we assume that parameter values are sorted so that
$\theta_1<\theta_2<\dots < \theta_K$ and all component magnitudes are no smaller than 1. 
We also assume that the minimum off-bound distance $\epsilon$ is sufficiently large and can be ignored during the proof.
    \par We
denote by $[l_j, u_j]$ the parameter range around each $\theta_j$ will be preserved after
thresholding with level $t$. We have $l_1 < \theta_1< u_1 < \dots < \theta_{j-1} < u_{j-1} < l_j <
\theta_j < u_j <  \dots < \theta_K$. Due to the important fact that the preserved parameter range
is small when the threshold level is large, as $t$ increase, $u _j$ decrease (i.e., the upper bound
decrease) and $l _j$ increase (i.e., the lower bound increase). So the proxy at $\omega=u_j$ has
value equal to the threshold, i.e.,
\begin{align}
\frac{t}{c_j} &= \frac{\nu \left( u_j \right)}{c_j} = \sum_{i=1}^K \frac{c_i}{c_j} \exp\left( - a \left|u_j - \theta_i \right| \right), \nonumber \\
\frac{t}{c_j} &= \sum_{i=1}^{j-1} \frac{c_i}{c_j} \exp\left( - a \left( u_j - \theta_i \right) \right) + \exp\left( - a (u_j - \theta_j) \right) + \sum_{i=j+1}^{K} \frac{c_i}{c_j} \exp\left( - a \left( \theta_i - u_j \right) \right), \nonumber \\
\frac{t}{c_j} &= \exp\left( - a \left( u_j-\theta_j \right) \right) \sum_{i=1}^{j-1} \frac{c_i}{c_j} \exp\left( - a \left(\theta_j - \theta_i \right) \right) + \exp\left( - a \left( u_j - \theta_j \right) \right), \nonumber \\
& \qquad + \exp(-a(\theta_j-u_j)) \sum_{i=j+1}^{K} \frac{c_i}{c_j} \exp\left( - a (\theta_i - \theta_j) \right), \nonumber \\
T_j & = A_j \frac{1}{U_j} + \frac{1}{U_j} + B_j U_j,
\label{equation_upper}
\end{align}
where $ U_j = \exp(a(u_j - \theta_j)) > 1 $,
\begin{align}
    0 \le &A_j = \sum_{i=1}^{j-1} \frac{c_i}{c_j} \exp\left( - a \left( \theta_j - \theta_i \right)
\right) \le r \sum_{i=1}^{j-1} \exp\left( - a \left(j - i\right) \zeta \right) \le r \frac{1-\exp\left(
- a \zeta j \right)}{1 - \exp\left( - a \zeta \right)} \le \frac{r}{\exp\left( a \zeta \right) - 1},
\\
0 \le &B_j = \sum_{i=j+1}^{K} \frac{c_i}{c_j} \exp\left( - a \left( \theta_i - \theta_j \right)
\right) \le r \sum_{i=1}^{K-j} \exp\left( - a \left(j - i \right) \zeta \right) \le r \frac{1-
\exp\left( - a \zeta (K-j+1) \right)}{1 - \exp\left( - a \zeta \right)} \nonumber \\
\le& \frac{r}{\exp\left( a
\zeta \right) - 1},
\end{align}
and $T_j = t/c_j$. One solution of the quadratic equation (\ref{equation_upper}) is
\begin{equation}
U_j = \frac{T_j-\sqrt{T_j^2 - 4(1+A_j)B_j}}{2B_j}.
\label{equation_uppersolution}
\end{equation}
In solution (\ref{equation_uppersolution}), $U_j$ decreases as $T_j$ increases. The alternative
solution is omitted due to the fact that $U_j$ will increase as $T_j$ increases, which implies that
$u _j$ will increase as $t$ decreases.
Similarly, 
\begin{align}
\frac{t}{c_j} &= \frac{\nu (l_j)}{c_j} = \sum_{i=1}^K \exp\left( - a \left| l_j - \theta_i \right| \right), \nonumber\\
\frac{t}{c_j} &= \sum_{i=1}^{j-1} \frac{c_i}{c_j} \exp\left( - a \left( l_j - \theta_i \right) \right) + \exp\left( - a \left( \theta_j-l_j \right) \right) + \sum_{i=j+1}^{K} \frac{c_i}{c_j} \exp\left( - a \left(\theta_i - l_j \right) \right), \nonumber\\
\frac{t}{c_j} &= \exp\left( -a \left( l_j-\theta_j \right) \right) \sum_{i=1}^{j-1} \frac{c_i}{c_j} \exp\left( -a \left( \theta_j - \theta_i \right) \right) + \exp\left( -a \left( \theta_j - l_j \right) \right), \nonumber \\
& \qquad + \exp\left( -a \left( \theta_j - l_j \right) \right) \sum_{i=j+1}^{K} \frac{c_i}{c_j} \exp\left( -a \left( \theta_i - \theta_j \right) \right), \nonumber \\
T_j & =  A_j L_j + \frac{1}{L_j} + B_j \frac{1}{L_j}, \label{equation_lower}
\end{align}
where $L_j = \exp\left(a\left( \theta_j - l_j\right)\right) > 1$. The corresponding solution is
\begin{equation}
L_j = \frac{T_j-\sqrt{T_j^2 - 4A_j(1+B_j)}}{2A_j}.
\label{equation_lowersolution}
\end{equation}\par

In order to have real values for solutions (\ref{equation_uppersolution}) and
(\ref{equation_lowersolution}), $T_j$ (and therefore $t$) should be sufficiently large. Since
\begin{align}
    \max \left( \left( 1 + A_j \right) B_j , A_j \left( 1 + B_j \right) \right) \le \left ( 1 +
\frac{r}{ \exp \left( a \zeta \right) - 1 } \right) \frac{ r }{ \exp \left( a \zeta \right) - 1 }
\le \frac{ 2 r^2 }{ \left( \exp \left( a \zeta \right) - 1 \right) ^2 },
\label{equation_both}
\end{align}
if
\begin{align} \label{equation:minThreshold}
T_j^2 = \left( \frac{t}{c_j} \right)^2 \ge \left( \frac{t}{c_{\max}} \right)^2 \ge \frac{ 8 r^2 }{
\left( \exp \left( a \zeta \right) - 1 \right) ^2 },
\end{align}
then both (\ref{equation_uppersolution}) and (\ref{equation_lowersolution}) are real valued.

Let $\hp_j \in \left[l_j, u_j \right]$ be the estimated parameter for $\theta_j$. Asymptotically,
when the sampling step of the parameter space $\Delta$ goes to zero, the balance weight properties
(\ref{equation:continuousbalance}) implies
\begin{equation}
\int_{l_j}^{\hp_j} \sum_{i=1}^K c_i \exp\left(-a \left| \omega - \theta_i \right| \right) \dw=
\int_{\hp_j}^{u_j} \sum_{i=1}^K c_i \exp\left(-a \left| \omega - \theta_i \right| \right) \dw.
\label{equation_balance}
\end{equation}
When $\hp_j \leq \theta_j$, the left hand side of (\ref{equation_balance}) is
\begin{align} \label{equation:temp_left}
& \frac{a}{c_j} \int_{l_j}^{\hp_j} \sum_{i=1}^K c_i \exp\left( -a \left| \omega - \theta_i \right|
\right) \dw \notag \\
= & a \sum_{i = 1}^{j-1} \frac{c_i}{c_j} \int_{l_j}^{\hp_j} \exp \left(-a \left( \omega - \theta_i
\right) \right) \dw + a \int_{l_j}^{\hp_j} \exp \left(-a \left(\theta_j - \omega \right) \right) \dw
+ a \sum_{i = j+1}^{K} \frac{c_i}{c_j} \int_{l_j}^{\hp_j} \exp \left(-a \left( \theta_i - \omega
\right) \right) \dw \notag \\
= & a A_j \int_{l_j}^{\hp_j} \exp \left( -a \left( \omega - \theta_j \right) \right) \dw + a
\int_{l_j}^{\hp_j} \exp \left( -a \left(\theta_j - \omega \right) \right) \dw + a B_j
\int_{l_j}^{\hp_j} \exp \left(-a \left( \theta_j - \omega \right) \right) \dw \notag \\
= & A_j \left( L_j - E_j \right) + \left( B_j+1\right) \left( \frac{1}{E_j} - \frac{1}{L_j} \right)
\notag \\
= & - A_j E_j + B_j \frac{1}{E_j} + \frac{1}{E_j} + A_j L_j - \frac{1}{L_j} - B_j \frac{1}{L_j},
\end{align}
where $E_j =1/\lambda(\hp_j - \theta_j) = \exp \left(a\left|\hp_j-\theta_j\right|\right) = \exp
\left( a \left( \theta_j-\hp_j \right) \right) \ge 1$, and the right hand side of
(\ref{equation_balance}) is 
\begin{align} \label{equation:temp_right}
& \frac{a}{c_j} \int_{\hp_j}^{u_j} \sum_{i=1}^K c_i \exp\left(-a \left| \omega - \theta_i \right|
\right) \dw \notag \\
= & a \sum_{i = 1}^{j-1} \frac{c_i}{c_j} \int_{\hp_j}^{u_j} \exp\left( -a \left( \omega - \theta_i
\right) \right) \dw + a \int_{\hp_j}^{u_j} \exp\left( -a \left|\omega -\theta_j \right| \right) \dw
+ a \sum_{i = j+1}^{K} \frac{c_i}{c_j} \int_{\hp_j}^{u_j} \exp\left( -a \left(\theta_i - \omega
\right) \right) \dw \notag \\
= & a \sum_{i = 1}^{j-1} \frac{c_i}{c_j} \int_{\hp_j}^{u_j} \exp\left( -a \left( \omega - \theta_i
\right) \right) \dw + a \int_{\hp_j}^{\theta_j} \exp\left( -a \left(\theta_j - \omega \right)
\right) \dw \notag \\
& \qquad + a \int_{\theta_j}^{u_j} \exp\left( -a \left( \omega -\theta_j \right) \right) \dw + a
\sum_{i = j+1}^{K} \frac{c_i}{c_j} \int_{\hp_j}^{u_j} \exp\left( -a \left(\theta_i - \omega \right)
\right) \dw \notag \\
= & A_j \left(E_j - \frac{1}{U_j} \right) + 1 - \frac{1}{E_j} + 1 - \frac{1}{U_j} + B_j \left( U_j-
\frac{1}{E_j} \right) \notag \\
= & A_j E_j - B_j \frac{1}{E_j} + 2 - \frac{1}{E_j} - A \frac{1}{U_j} - \frac{1}{U_j} + B_j U_j.
\end{align}
After plugging (\ref{equation:temp_left}) and (\ref{equation:temp_right}) into (\ref{equation_balance}) and moving all terms with $L_j$ or $U_j$ to the right side and moving other terms to the left side, we obtain
\begin{equation}
\begin{aligned}
    & 2 A_j E_j - 2 B_j \frac{1}{E_j} + 2 - \frac{2}{E_j}\\
= & A_j L_j - \frac{1}{L_j} - B_j \frac{1}{L_j} + A_j \frac{1}{U_j} + \frac{1}{U_j} - B_j U_j\\
= & A_j L_j + A_j L_j - T_j + T_j - B_j U_j - B_j U_j\\
= & 2 A_j L_j - 2 B_j U_j.\\
= & \left( T_j-\sqrt{T_j^2 - 4A_j(1+B_j)} \right) - \left( T_j - \sqrt{T_j^2 - 4(1+A_j)B_j} \right)\\
= & \sqrt{T_j^2 - 4(1+A_j)B_j} - \sqrt{T_j^2 - 4A_j(1+B_j)}\\
= & \frac{ \left( T_j^2 - 4(1+A_j)B_j \right) - \left( T_j^2 - 4A_j(1+B_j) \right) }{\sqrt{T_j^2 - 4A_j(1+B_j)} + \sqrt{T_j^2 - 4(1+A_j)B_j}}\\
= & \frac{4 \left( A_j - B_ j \right) }{\sqrt{T_j^2 - 4A_j(1+B_j)} + \sqrt{T_j^2 - 4(1+A_j)B_j}}
\end{aligned},
\end{equation}
where the second equality results from (\ref{equation_upper}) and (\ref{equation_lower}).

One can show a similar result when $\theta_j > \hp_j$ and $E_j = \exp\left( a \left| \theta_j -
\hp_j \right| \right) = \exp\left( a \left( \theta_j - \hp_j \right) \right)$, so
(\ref{equation_balance}) can be written as
\begin{equation}
\frac{ A_j - B_j}{S_j} = \left\{ \begin{aligned}
A_j E_j - B_j \frac{1}{E_j} + 1 - \frac{1}{E_j} &\qquad \text{if} \quad \theta_j \le \hp_j\\
A_j \frac{1}{E_j} - B_j E_j - 1 + \frac{1}{E_j} &\qquad \text{if} \quad \theta_j > \hp_j\\
\end{aligned} \right.,
\label{equation:simplebalance}
\end{equation}
where 
\begin{equation}
S_j =\frac{1}{2} \left( \sqrt{T_j^2 - 4A_j(1+B_j)} + \sqrt{T_j^2 - 4(1+A_j)B_j} \right) \le T_j \le
\frac{t}{c_{\min}} \le 1 \le E_j,
\end{equation}
and
\begin{equation}
S_j =\frac{1}{2} \left( \sqrt{T_j^2 - 4A_j(1+B_j)} + \sqrt{T_j^2 - 4(1+A_j)B_j} \right) \ge
\sqrt{\left( \frac{t}{r c_{\min}} \right) - 8 \left( \frac{r}{\exp\left( a \zeta \right) - 1}
\right)^2 }
\end{equation}
using (\ref{equation_both}).

When $\theta_j \le \hp_j$, 
\begin{equation}
(A_j - B_j) / S_j = A_j E_j - B_j / E_j + 1 - 1 / E_j \ge ( A_j - B_j ) / E_j,
\end{equation} 
So $A_j \ge B_j$ due to the fact that $S_j \le E_j$. If
\begin{equation}
S_j \ge \sqrt{\left( \frac{t}{r c_{\min}} \right) - 8 \left( \frac{r}{\exp\left( a \zeta \right) -
1} \right)^2 } \ge \exp(-a\sigma) \ge 0,
\end{equation}
which can be rewritten as
\begin{equation} \label{equation:minSeparation}
\zeta \geq \frac{1}{a} \ln \left( \sqrt{ \frac{8 r^2}{ t^2/\left(r c_{\min} \right)^2 -
\exp(-2a\sigma) } }+1 \right),
\end{equation}
the condition (\ref{equation:minThreshold}) satisfies, so the solutions
(\ref{equation_uppersolution}) and (\ref{equation_lowersolution}) are both real valued.
Additionally, we have
\begin{equation}
A_j E_j - B_j \frac{1}{E_j} + 1 - \frac{1}{E_j} = \frac{ A_j - B_j}{S_j} \le \left( A_j - B_j
\right) \exp(a \sigma) \le A_j \exp(a \sigma) - B_j \exp(- a \sigma) + 1 - \exp( - a \sigma).
\end{equation}
This implies that $E_j \le \exp\left( a \sigma \right)$ and hence $0 \le \hp_j - \theta_j \le
\sigma$. Similarly, when $\theta_j > \hp_j$, $A_j \le B_j$ and $- \sigma \le \hp_j - \theta_j \le
0$. 

Finally, the condition for $\zeta$ to be real valued is $t \ge r c_{\min} \exp (-a \sigma)$, which can be
obtained from (\ref{equation:minSeparation}).

\section*{References}

\bibliographystyle{elsarticle-num}
\biboptions{sort&compress}
\bibliography{bibliography}

\end{document}